\documentclass[aip,reprint]{revtex4-2}
\usepackage{graphicx} % Required for inserting images
\usepackage{amsmath,amsfonts,amssymb} 
\usepackage{hyperref}
\usepackage{comment}
\usepackage[utf8]{inputenc}
\usepackage{cleveref}
\usepackage{lipsum}
\usepackage{mathtools}

\begin{document}
\title{The neutron spectral moments method in the time-of-flight domain}
\author{Brian D. Appelbe}
\email{bappelbe@ic.ac.uk}
\author{Aidan J. Crilly} 
\affiliation{Centre for Inertial Fusion Studies, The Blackett Laboratory, Imperial College, London SW7 2AZ, United Kingdom}

\begin{abstract}
An analysis methodology is developed for the time-of-flight (TOF) signals recorded by two or more collinear neutron detectors located at different distances from a pulsed neutron source. It is based on taking central moments of the TOF signals and relating these to a set of co-moments of the distribution of production times and velocities of neutrons emitted towards the detectors. Given $n$ detectors, we can obtain all such co-moments of order $n-1$ and lower. These co-moments contain information on the time-varying behaviour of the neutron source. A physical interpretation is provided for several co-moments of interest. 
\end{abstract}

\maketitle
%%%%%%%%%%%%%%%%%%%%%%%%%%%%%%%%%%%%%%%%%%%%%%%%%%%%%%%%%%%%%%%%%%%%%%%%%%%%%%%%%%%%%%%
%%%%%%%%%%%%%%%%%%%%%%%%%%%%%%%%%%%%%%%%%%%%%%%%%%%%%%%%%%%%%%%%%%%%%%%%%%%%%%%%%%%%%%%
%%%%%%%%%%%%%%%%%%%%%%%%%%%%%%%%%%%%%%%%%%%%%%%%%%%%%%%%%%%%%%%%%%%%%%%%%%%%%%%%%%%%%%%
%%%%%%%%%%%%%%%%%%%%%%%%%%%%%%%%%%%%%%%%%%%%%%%%%%%%%%%%%%%%%%%%%%%%%%%%%%%%%%%%%%%%%%%
%%%%%%%%%%%%%%%%%%%%%%%%%%%%%%%%%%%%%%%%%%%%%%%%%%%%%%%%%%%%%%%%%%%%%%%%%%%%%%%%%%%%%%%
\begin{comment}
\section{To Do:}
\begin{itemize}
    \item Check all derivations
    \item Validity of expansion before integration?
    \item a set of generalized TOF moments?
\end{itemize}    
\end{comment}

\section{Introduction}\label{s:0}
Neutron time-of-flight (nTOF) detectors are used in Inertial Confinement Fusion (ICF) and other pulsed neutron source experiments to obtain information on the emitted neutrons. Physical models can be used to infer conditions of the neutron emitting plasma from this information. The basic principle of nTOF detectors is that they record the flux as a function of time of neutrons arriving at a detector which is located at a known distance from the source, a quantity that we will refer to as the \textit{time-of-flight} (TOF)  signal. The arrival time of a neutron is a function of the time at which it is created in the source, the \textit{production time}, and its velocity. Thus, nTOF detectors can provide information on the time-varying behaviour of the neutron source. The distribution of velocities and production times of neutrons reaching a detector will be referred to as the \textit{neutron source function} (NSF).

Accurate identification of the NSF has been a longstanding challenge. This is due to a degeneracy between neutron velocity and production time: a slow neutron produced earlier in time can arrive coincidentally with a faster neutron produced later. In the case of ICF the short duration of neutron pulses and large neutron yields have provided a means to avoid this degeneracy: nTOF detectors are placed at a sufficiently large distance from the detector such the arrival time is dominated by neutron velocity and variations in production time can be neglected.\cite{Moore_RSI2023} In effect, one assumes that all neutron production is instantaneous or, equivalently, that the arrival times provide a record of neutron velocities irrespective of their production time (a time-integrated neutron velocity spectrum). This luxury is not available to those working with other pulsed neutron sources.

The ability to measure a time-integrated neutron velocity spectrum, coupled with the stringent requirements placed on diagnostics in ICF experiments,\cite{Kilkenny_POP2024} inspired the development of a rigorous theory for interpreting the spectra of velocities of neutrons produced by deuterium-deuterium (DD) and deuterium-tritium (DT) reactions which do not undergo any subsequent scattering (often referred to as the DD and DT primary spectra or primary peaks). This theory involves calculating, or "taking", moments of the primary spectra in the neutron velocity domain and using the reaction cross section and basic kinematics to relate these moments to certain averagings over the ion distribution functions of the neutron-emitting plasma. For ICF experiments, the moments method has provided a means for inferring quantities such as ion temperature and hotspot velocity, and identifying the presence of non-Maxwellian ion distributions in the plasma. The authoritative work on the moments method is that of Munro\cite{Munro_2016} for Maxwellian ion distributions, while Crilly et al\cite{Crilly_2022} extended it to non-Maxwellian ion distributions. Appelbe et al\cite{APPELBE_HEDP2024} made a recent attempt to innovate the work of Munro and proposed an alternative set of moments, although the results remain untested in the laboratory.

In the present work we explore how the neutron spectral moments method can be applied in the TOF domain, that is, to the record of neutron arrival times on an nTOF detector. If one can assume that the neutron source is instantaneous then this is a just a trivial extension of the moments method in the neutron velocity domain (there is a one-to-one relationship between neutron velocity and arrival time). However, when this assumption is invalid, perhaps because the nTOF detector is not located sufficiently far from the source, the degeneracy issue cannot be avoided. Our approach to tackling this degeneracy and gaining a physical understanding of the moments of TOF signals can be summarised by the following key ideas
\begin{enumerate}
    \item \textit{Central moments} of the TOF signal on a detector at a given distance from the source are equivalent to a summation 
    of \textit{inverse-velocity co-moments} of the NSF. These inverse-velocity co-moments are defined and examined in detail in section \ref{s:3}.
    \item By taking central moments of the TOF signal on detectors at multiple different distances we can solve for the values of individual inverse-velocity co-moments of the NSF. The polynomial equations \eqref{e:2.1a} and \eqref{e:2.1b} govern the relationship between central moments of the TOF and inverse-velocity co-moments of the NSF.
    \item A physical interpretation of the inverse-velocity co-moments of the NSF gives information on the time-varying behaviour of ion distribution functions. Some examples of this are given in \ref{s:3.4}, for arbitrary ion distribution functions, and in \ref{s:4}, for Maxwellians.
\end{enumerate}
A key requirement for the second point above is that the detectors are \textit{collinear}, that is, they lie along a single direction of neutron emission with respect to the neutron source. There are many physical processes which can cause neutron emission from a source to be anisotropic and, consequently, a NSF that varies with neutron emission direction. Thus, it is only when detectors are collinear that we can be assured that TOF signals at different distances are a function of the same NSF. Other requirements and assumptions which underpin our work include
\begin{itemize}
    \item Detectors are co-timed. This assumption is required so that we can solve for the inverse-velocity co-moments given central moments of the TOF on multiple detectors at different distances.
    \item Neutrons are emitted from a point source. Our analysis requires that all neutrons traverse the same distance to reach a detector.
    \item We assume that an ideal nTOF detector records the TOF signal. In practice, issues such as detector resolution, noise, etc may limit the accuracy of measurement of the TOF signal. Given this assumption, it is worthwhile to mention that a range of detector concepts exist which could record the TOF signal. In addition to conventional nTOF detectors,\cite{Moore_RSI2023} concepts in which neutrons are converted to detectable gamma rays\cite{Meaney_2022} or induce the Pockels effect in a suitable crystal\cite{Arikawa_RSI2020} at certain distances from the source have been proposed. Furthermore, diagnostics such as the GRH diagnostic,\cite{Hermann_RSI2010,Geppert-Kleinrath_RSI2018} which records the flux of gamma rays emitted from DT reactions as a function of time can be treated as TOF detector at zero distance from the source under certain assumptions (namely, that the DT branching ratio is known and other sources of gammas are neligible). Finally, it is important to mention that detector concepts that would provide a direct measurement of the temporal history of neutron emission have also been proposed.\cite{Frenje_RSI2016,Kunimune_RSI2022} Such detectors would not require the analysis that we construct in this work.
    \item We assume that environmental neutron scattering from sources such as target chambers, hardware, beamline attenuation, etc is negligible.
    \item In sections \ref{s:3} and \ref{s:4}, where we offer a physical interpretation of the inverse-velocity co-moments of the NSF, we consider only the primary, unscattered neutron spectra from DD and DT reactions. We assume that no neutron scattering occurs in the source and neutron velocities depend only on reaction kinematics and the Q-value of the reaction.
    \item We use non-relativistic kinematics throughout. While a fully relativistic treatment may introduce some minor corrections,\cite{Munro_2016,APPELBE_2014} our methodology and conclusions should not be affected.
\end{itemize}

Our application of the moments method to the TOF domain is novel. Nevertheless, there is an extensive history of analyses of TOF signals. The original work appears to be that of Vlad,\cite{Vlad_1984,Vlad_1989} in which numerical solution of an integral equation relating the TOF signal and NSF was proposed. Schmidt and Herold proposed solving this equation using a Laplace transform\cite{Schmidt_1987}, while several groups \cite{Tiseanu_1996a,Tiseanu_1996b,Turcanu_2002,Denisova_2004,Rezac_2011,Rezac_2012,Catenacci_2020} have investigated statistical methods including Monte Carlo reconstruction, genetic algorithms and maximum a posteriori algorithms. The common feature of all these methods is that they have been attempting to reconstruct the NSF from measurement of the TOF signal. Such reconstructions are always susceptible to the degeneracy issue. The moments method that we present here differs in that we do not seek to reconstruct the NSF. Instead we ask \textit{what information can be deterministically obtained from TOF signals measured at different distances?} The inverse-velocity co-moments are this information.

Finally, we note that there has been some recent work in the design of collinear time-of-flight detectors for NIF and other ICF facilities, which have motivated our present work.\cite{Moore_2022, Schlossberg_RSI2022, Meaney_2022} Computational models of TOF signals suggested that such a detector system could identify the presence of a time-varying ion temperature in the hotspot, while experiments with two collinear nTOF detectors at different distances\cite{Mitrani_RSI2024,Grim_2024} have been used to infer values of bang time and hotspot velocity. The theory that we develop in our work here provides a validation of these observations.

\subsection{Outline of Contents}\label{s:0.2}
We begin in section \ref{s:1} by defining basic quantities and relationships that will be used in the work. Section \ref{s:2} shows how central moments of the TOF signal are a summation of inverse-velocity co-moments of the NSF and how multiple detectors at different distances allows us to solve for individual inverse-velocity co-moments. Sections \ref{s:3} and \ref{s:4} are concerned with the physical interpretation of these inverse-velocity co-moments, first by assuming arbitrary ion velocity distributions (section \ref{s:3}) and then for the particular case of Maxwellian ion velocity distributions (section \ref{s:4}). Several ancillary results are presented in section \ref{s:DA}, including a model for choosing the detector distances in subsection \ref{s:A2}. It is shown that these distances will depend on the neutron source duration and the range of neutron velocities produced.

%%%%%%%%%%%%%%%%%%%%%%%%%%%%%%%%%%%%%%%%%%%%%%%%%%%%%%%%%%%%%%%%%%%%%%%%%%%%%%%%%%%%%%%
%%%%%%%%%%%%%%%%%%%%%%%%%%%%%%%%%%%%%%%%%%%%%%%%%%%%%%%%%%%%%%%%%%%%%%%%%%%%%%%%%%%%%%%
%%%%%%%%%%%%%%%%%%%%%%%%%%%%%%%%%%%%%%%%%%%%%%%%%%%%%%%%%%%%%%%%%%%%%%%%%%%%%%%%%%%%%%%
%%%%%%%%%%%%%%%%%%%%%%%%%%%%%%%%%%%%%%%%%%%%%%%%%%%%%%%%%%%%%%%%%%%%%%%%%%%%%%%%%%%%%%%
%%%%%%%%%%%%%%%%%%%%%%%%%%%%%%%%%%%%%%%%%%%%%%%%%%%%%%%%%%%%%%%%%%%%%%%%%%%%%%%%%%%%%%%

\section{Basic Concepts}\label{s:1}
Consider a neutron which is produced at time $t$ and with a velocity $v$. This neutron will arrive at a detector located a distance $x$ from the neutron source at time $\tau$, given by
\begin{equation}\label{e:1.1}
    \tau = t+\frac{x}{v}.
\end{equation}
The TOF will be denoted by $S\left(x,\tau\right)$ and the NSF by $f\left(t,v\right)$. The velocity $v$ in \eqref{e:1.1} and in the NSF is not a vector. It is the magnitude of the velocity of neutrons that are emitted in the direction of the detector.

%%%%%%%%%%%%%%%%%%%%%%%%%%%%%%%%%%%%%%%%%%%%%%%%%%%%%%%%%%%%%%%%%%%%%%%%%%%%%%%%%%%%%%%
%%%%%%%%%%%%%%%%%%%%%%%%%%%%%%%%%%%%%%%%%%%%%%%%%%%%%%%%%%%%%%%%%%%%%%%%%%%%%%%%%%%%%%%
%%%%%%%%%%%%%%%%%%%%%%%%%%%%%%%%%%%%%%%%%%%%%%%%%%%%%%%%%%%%%%%%%%%%%%%%%%%%%%%%%%%%%%%

\subsection{Relationship between TOF and NSF}\label{s:1.1}
The TOF can be defined in terms of the NSF as
\begin{equation}\label{e:1.1.1}
    S\left(x,\tau\right) = \int\delta\left(\tau-t-xv^{-1}\right)f\left(t,v\right)dvdt,
\end{equation}
where the integration sign denotes integration over the entire domains of $t$ and $v$, $-\infty <t<\infty$ and $0<v<\infty$. (Throughout this work an integral can be assumed to be over the entire domain indicated by the differentials, unless otherwise stated). From this definition, we can see that the TOF consists of integration along \textit{curves of projection} in the $\left(t,v\right)$ space. That is, for a given $x$ and $\tau$, \eqref{e:1.1} defines a curve in $\left(t,v\right)$ space. All neutrons that are produced on a given curve of projection will have the same arrival time $\tau$ at a detector located at $x$.

We can integrate over the Dirac delta function in \eqref{e:1.1.1} to make the relationship between TOF, NSF and curves of projection more explicit
\begin{equation}\label{e:1.1.2}
    S\left(x,\tau\right) = \int f\left(\tau-\frac{x}{v},v\right)dv \equiv \int f\left(t,\frac{x}{\tau-t}\right)dt.\nonumber
\end{equation}

Curves of projection are an important concept as they provide the link between the TOF and NSF. They also form the foundation for "tomographic reconstruction" techniques for estimating the NSF given multiple measurements of the TOF at different distances. We do not investigate such techniques here, but we do use curves of projection for choosing the distances to detectors, as described in section \ref{s:A2}.

%%%%%%%%%%%%%%%%%%%%%%%%%%%%%%%%%%%%%%%%%%%%%%%%%%%%%%%%%%%%%%%%%%%%%%%%%%%%%%%%%%%%%%%
%%%%%%%%%%%%%%%%%%%%%%%%%%%%%%%%%%%%%%%%%%%%%%%%%%%%%%%%%%%%%%%%%%%%%%%%%%%%%%%%%%%%%%%
%%%%%%%%%%%%%%%%%%%%%%%%%%%%%%%%%%%%%%%%%%%%%%%%%%%%%%%%%%%%%%%%%%%%%%%%%%%%%%%%%%%%%%%

\subsection{Taking moments of TOF}\label{s:1.2}
Given any function of $\tau$, denoted by $g\left(\tau\right)$, we can calculate a moment of the TOF with respect to $\tau$. This moment is denoted by $\langle g\rangle$ and is defined as
\begin{equation}\label{e:1.2.1}
    \langle g\rangle = \int g\left(\tau\right)S\left(x,\tau\right)d\tau,\nonumber
\end{equation}
where integration is over the entire domain of $\tau$, $-\infty<\tau<\infty$. The moment $\langle g\rangle$ is a function of $x$ only. We can use \eqref{e:1.1.1} to relate moments of TOF to co-moments of the NSF
\begin{eqnarray}\label{e:1.2.2}
    \langle g\rangle &=& \int g\left(\tau\right)S\left(x,\tau\right)d\tau\nonumber\\
    &\equiv& \int g\left(t+\frac{x}{v}\right)f\left(t,v\right)dvdt.
\end{eqnarray}
Here, we integrate over $t$ and $v$ on the right-hand side, and so $\langle g\rangle$ may be thought of as both a co-moment of the NSF with respect to $t$ and $v$ and a moment of the TOF with respect to $\tau$.

The moment operator $\left\langle\right\rangle$ is a linear operator, obeying 
\begin{equation}\label{en:1}
    \left\langle c\left(g+h\right)\right\rangle =  c\left\langle g\right\rangle+ c\left\langle h\right\rangle, 
\end{equation}
where $c$ is a constant. We will make much use of this linearity property. We will also introduce various other moment operators that will obey this property. These operators will be denoted by $\left\langle\right\rangle$ with an appropriate subscript.
%%%%%%%%%%%%%%%%%%%%%%%%%%%%%%%%%%%%%%%%%%%%%%%%%%%%%%%%%%%%%%%%%%%%%%%%%%%%%%%%%%%%%%%
%%%%%%%%%%%%%%%%%%%%%%%%%%%%%%%%%%%%%%%%%%%%%%%%%%%%%%%%%%%%%%%%%%%%%%%%%%%%%%%%%%%%%%%
%%%%%%%%%%%%%%%%%%%%%%%%%%%%%%%%%%%%%%%%%%%%%%%%%%%%%%%%%%%%%%%%%%%%%%%%%%%%%%%%%%%%%%%

\subsection{Normalizing moments of TOF}\label{s:1.3}
Both $S\left(x,\tau\right)$ and $f\left(t,v\right)$ are un-normalized quantities. They have dimensions of (\textit{neutrons arriving}) per unit time per unit solid angle and (\textit{neutrons produced}) per unit time per unit solid angle per unit velocity magnitude, respectively. The total number of neutrons per unit solid angle is defined as
\begin{equation}\label{e:1.3.1}
    \langle 1\rangle = \int S\left(x,\tau\right)d\tau = \int f\left(t,v\right)dvdt.\nonumber
\end{equation}
The notation $\langle 1\rangle$ indicates that it can be considered as a moment (co-moment) of the TOF (NSF). This quantity is independent of the detector distance, since the function defining the moment, $g$, is independent of $x$. It can be used to define normalized moments of the TOF and NSF, denoted by $\langle g\rangle_{0}$,
\begin{equation}\label{e:1.3.2}
    \langle g\rangle_{0} = \frac{\langle g\rangle}{\langle 1\rangle} = \int g\left(\tau\right)\frac{S\left(x,\tau\right)}{\langle 1\rangle}d\tau.
\end{equation}
Equation \eqref{e:1.3.2} shows that normalized moments are equivalent to moments of the normalized TOF. There are several advantages to dealing with normalized moments $\langle g\rangle_{0}$ instead of un-normalized moment $\langle g\rangle$. First, using a normalized TOF removes some challenges relating to detector sensitivity, which is worth considering even though we are assuming an ideal detector in the present work. Secondly, since the normalizing quantity $\langle 1\rangle$ is the number of neutrons per unit solid angle emitted in a particular direction, any physical quantities inferred from $\langle g\rangle_{0}$ will be \textit{directed yield-averaged quantities}, that is, the physical quantities inferred will be independent of the number of neutrons emitted towards the detectors. As indicated by the use of the term directed, these physical quantities can vary with the direction of neutron emission. Physical interpretations of $\langle g\rangle_{0}$ will be further discussed in sections \ref{s:4} and \ref{s:5}.

%%%%%%%%%%%%%%%%%%%%%%%%%%%%%%%%%%%%%%%%%%%%%%%%%%%%%%%%%%%%%%%%%%%%%%%%%%%%%%%%%%%%%%%
%%%%%%%%%%%%%%%%%%%%%%%%%%%%%%%%%%%%%%%%%%%%%%%%%%%%%%%%%%%%%%%%%%%%%%%%%%%%%%%%%%%%%%%
%%%%%%%%%%%%%%%%%%%%%%%%%%%%%%%%%%%%%%%%%%%%%%%%%%%%%%%%%%%%%%%%%%%%%%%%%%%%%%%%%%%%%%%

\subsection{Primary spectra and ion distribution functions}\label{s:1.4}
In our discussion of TOF and NSF so far we have made no assumption about the physical mechanisms occurring in the neutron emitting plasma that create the NSF. Relationships between moments of the TOF and co-moments of the NSF, defined by \eqref{e:1.2.2}, can always be established regardless of the mechanism. However, in this work we are particularly interested in the primary neutron spectra produced by DD and DT reactions. We assume that these neutrons do not undergo any scattering and so the NSF depends only on the distribution functions of the reacting ions, the Q-value of the reaction and its differential cross section. The relationship between NSF and ion distribution functions has been thoroughly investigated in the literature,\cite{APPELBE_HEDP2024,Crilly_2022,Munro_2016,APPELBE_2014,Appelbe_2011} and so we simply summarise some relevant details here.

The ion distribution functions are defined by the joint distribution of centre-of-mass (CM) velocity, $\vec{v_{c}}$, and relative velocity, $\vec{v_{r}}$, of reacting ion pairs.\cite{APPELBE_HEDP2024,Crilly_2022} This joint distribution is denoted by $F\left(t,\vec{v_{c}},\vec{v_{r}}\right)$ and is given by
\begin{equation}\label{en:2}    F\left(t,\vec{v_{c}},\vec{v_{r}}\right)d^{3}\vec{v_{c}}d^{3}\vec{v_{r}} = f_{1}\left(t,\vec{v_{1}}\right)f_{2}\left(t,\vec{v_{2}}\right)d^{3}\vec{v_{1}}d^{3}\vec{v_{2}},\nonumber
\end{equation}
where $f_{1}\left(t,\vec{v_{1}}\right)d^{3}\vec{v_{1}}$ and $f_{2}\left(t,\vec{v_{2}}\right)d^{3}\vec{v_{2}}$ are the velocity distributions of the two reacting species. The joint distribution can vary in both time and space. We use the production time variable $t$ to denote temporal dependence. For brevity, we omit explicit notation of the spatial variation, and use $dV$ to indicate integration over the spatial domain. Furthermore, we assume that the neutron source size is sufficiently small to be treated as a point source. This ensures that all neutrons travel the same distance $x$ to reach a detector, a necessary assumption for our analysis.

Given an ion distribution function, we can calculate the NSF using
\begin{equation}\label{e:1.4.1}
        f\left(t,v\right) = \int\delta\left(\xi_{cr}-v\right) R_{k}d^{3}\vec{v_{c}}d^{3}\vec{v_{r}}dV,
\end{equation}
where
\begin{eqnarray}
    v &\equiv& \xi_{cr} = \vec{v_{c}}\cdot\hat{v}+\sqrt{u^{2}-v_{c}^{2}+\left(\vec{v_{c}}\cdot\hat{v}\right)^{2}},\label{e:1.4.2a}\\
    R_{k} &=& \frac{v^{2}}{u\sqrt{u^{2}-v_{c}^{2}+\left(\vec{v_{c}}\cdot\hat{v}\right)^{2}}}v_{r}\frac{d\sigma}{d\Omega_{cm}}F\left(t,\vec{v_{c}},\vec{v_{r}}\right).\label{e:1.4.2b}
\end{eqnarray}
In the above, $v$ denotes the magnitude neutron velocity when it is treated as an independent variable and $\xi_{cr}$ denotes the same quantity when it is treated as a function of relative velocity and CM velocity vectors.\cite{APPELBE_HEDP2024} The NSF depends on the direction of neutron emission which is defined by the term $\hat{v}$. This denotes a unit vector directed from the neutron source towards the detector, and so, for example, $\vec{v_{c}}\cdot\hat{v}$ is the component of CM velocity directed towards the detectors. Also in the above equations, $u$ is the magnitude of neutron velocity in the CM frame (defined in appendix \ref{s:A1.1}), $R_{k}$ is referred to as the reactivity kernel and $\frac{d\sigma}{d\Omega_{cm}}$ is the differential cross section for the reaction. As discussed in appendix \ref{s:A1.2}, the first term in $R_{k}$ is a Jacobian determinant arising from a frame transformation of the differential cross section. Approximations of this term will play an important role in section \ref{s:3} when we seek a physical interpretation of co-moments of the NSF.

Now, combining \eqref{e:1.2.2} and \eqref{e:1.4.1} gives the following relation between moments of TOF and integration over the distribution functions
\begin{equation}
    \langle g\rangle = \int g\left(t+\frac{x}{\xi_{cr}}\right)R_{k}d^{3}\vec{v_{c}}d^{3}\vec{v_{r}}dVdt,\label{e:1.4.3}
\end{equation}
This expression demonstrates that moments of the TOF, which we already know are equivalent to co-moments of the NSF, are equivalent to co-moments of the ion distribution functions with respect $t$, $\vec{v_{c}}$ and $\vec{v_{r}}$. Establishing relationships between moments of the TOF and co-moments of the ion distribution functions is the focus of this work. Co-moments of the NSF provide an important intermediate step in this process.

Finally, we note that we can also use \eqref{e:1.4.1} to define a more general set of co-moments of the NSF than those defined by \eqref{e:1.2.2}. That is, for any function of $t$ and $v$, denoted by $g\left(t,v\right)$, we have
\begin{eqnarray}\label{e:1.4.4}
    \langle g\left(t,v\right)\rangle_{0} &\equiv& \frac{\left\langle g\left(t,v\right)\right\rangle}{\left\langle 1\right\rangle}\nonumber\\
    &=& \frac{1}{\langle 1\rangle}\int g\left(t,\xi_{cr}\right)R_{k}d^{3}\vec{v_{c}}d^{3}\vec{v_{r}}dVdt.
\end{eqnarray}
In contrast, the co-moments defined in \eqref{e:1.2.2} are for a restricted set of functions of $t$ and $v$, obeying \eqref{e:1.1}. Of course, $\langle g\left(t,v\right)\rangle_{0}$ cannot be equated to moments of the TOF signal (the observables in an experiment) unless the variables $t$ and $v$ obey \eqref{e:1.1}, such that $g\left(t,v\right)=g\left(\tau\right)$.

%%%%%%%%%%%%%%%%%%%%%%%%%%%%%%%%%%%%%%%%%%%%%%%%%%%%%%%%%%%%%%%%%%%%%%%%%%%%%%%%%%%%%%%
%%%%%%%%%%%%%%%%%%%%%%%%%%%%%%%%%%%%%%%%%%%%%%%%%%%%%%%%%%%%%%%%%%%%%%%%%%%%%%%%%%%%%%%
%%%%%%%%%%%%%%%%%%%%%%%%%%%%%%%%%%%%%%%%%%%%%%%%%%%%%%%%%%%%%%%%%%%%%%%%%%%%%%%%%%%%%%%
%%%%%%%%%%%%%%%%%%%%%%%%%%%%%%%%%%%%%%%%%%%%%%%%%%%%%%%%%%%%%%%%%%%%%%%%%%%%%%%%%%%%%%%
%%%%%%%%%%%%%%%%%%%%%%%%%%%%%%%%%%%%%%%%%%%%%%%%%%%%%%%%%%%%%%%%%%%%%%%%%%%%%%%%%%%%%%%

\section{Central moments of TOF}\label{s:2}
Given our definition of moments in \eqref{e:1.2.2}, we have freedom to choose whatever function $g\left(\tau\right)$ which will provide the maximum information about the NSF or the ion distribution functions. However, the process of selecting such a function is non-trivial. Appelbe et al\cite{APPELBE_HEDP2024} identified an interesting set of functions when working in the neutron velocity domain (a result that we will return to in section \ref{s:6}), but it is not yet clear that an analogous set exists in the time-of-flight domain. Therefore, for this work we resort to a conventional choice for $g\left(\tau\right)$, namely the set of functions that will give us \textit{central moments} of TOF. These functions are
\begin{equation}
    g\left(\tau\right)=\begin{cases}
			\tau, & \text{if $n = 1$}\\
            \left(\tau-\langle \tau\rangle_{0}\right)^{n}, & \text{if $n>1$, $n\in\mathbb{N}$}
		 \end{cases}\nonumber
\end{equation}
where $\mathbb{N}$ is the set of natural numbers excluding $0$. We refer to $n$ as the order of the central moment. The set of central moments of the TOF is related to co-moments of the NSF as follows
\begin{eqnarray}
    &&\langle\tau\rangle_{0} = \langle t\rangle_{0}+x\left\langle v^{-1}\right\rangle_{0},\label{e:2.1a}\\
    &&\left\langle\left(\tau-\langle \tau\rangle_{0}\right)^{n}\right\rangle_{0} =\sum_{k=0}^{n}x^{k}\binom{n}{k}\nonumber\\
    &&\times\left\langle\left(t-\langle t\rangle_{0}\right)^{n-k}\left( v^{-1}-\left\langle v^{-1}\right\rangle_{0}\right)^{k}\right\rangle_{0},\quad n>1.\label{e:2.1b}
\end{eqnarray}
From these equations we see that the $n$-th order moment of the TOF is equivalent to a polynomial in $x$ of degree $n$ whose  coefficients are co-moments of the NSF multiplied by a binomial coefficient. We call these co-moments the \textit{inverse-velocity co-moments} to emphasize that it is powers of the inverse of neutron velocity which appear in the co-moment expressions. The inverse-velocity co-moments can be related to ion distribution functions, using \eqref{e:1.4.4}. The order of an inverse-velocity co-moment is the sum of the orders of the time and inverse-velocity components of the co-moment, that is all co-moments on the right-hand side of \eqref{e:2.1b} are of order $n$.

The importance of \eqref{e:2.1a} and \eqref{e:2.1b} is that they illustrate how measurements of TOF on collinear detectors can be used to calculate the values of inverse-velocity co-moments of the NSF, a key result of this work. \textit{Assume that we have $n$ collinear detectors at different distances $x$ from the neutron source. By calculating the moment of order $n-1$ of the TOF on every detector we will be able to calculate the coefficients of the polynomial in $x$ of degree $n-1$.} Therefore, we can calculate all inverse-velocity co-moments of order $n-1$ and lower given $n$ detectors by solving a system of linear equations.

In the remainder of this section we will analyze the moments of TOF and the inverse-velocity co-moments of orders to $1$ to $3$ to illustrate the information that can be obtained from measurements of the TOF, but first we introduce a succinct notation for the inverse-velocity co-moments that we will use for the remainder of this work. The notation is
\begin{equation}
    C\left(t,v^{-1}\right)^{j,k}_{0} = \left\langle\left(t-\left\langle t\right\rangle_{0}\right)^{j}\left(v^{-1}-\left\langle v^{-1}\right\rangle_{0}\right)^{k}\right\rangle_{0},\label{e:4.9a}
\end{equation}
for $j,k\in\mathbb{N}_{0}$, $j+k>1$, where $\mathbb{N}_{0}$ is the set of natural numbers including $0$. With this notation \eqref{e:2.1b} becomes
\begin{eqnarray}
    C\left(\tau\right)^{n}_{0} &=& \sum_{k=0}^{n}x^{k}\binom{n}{k}C\left(t,v^{-1}\right)^{n-k,k}_{0},\quad n>1.\label{e:2.0b}
\end{eqnarray}
As can be seen from the left-hand side of \eqref{e:2.0b}, the notation can equally be applied to the moment of TOF. The number of superscripts identifies whether a moment ($1$ superscript) or co-moment ($2$ superscripts) is under consideration. Some basic relations that we will make use of later are
\begin{eqnarray}
   C\left(a,b\right)^{1,1}_{0} &=& \left\langle ab\right\rangle_{0} -  \left\langle a\right\rangle_{0} \left\langle b\right\rangle_{0},\nonumber\label{e:4.9c}\\
   C\left(a,b\right)^{2,1}_{0} &=&  \left\langle a^{2}b\right\rangle_{0}- \left\langle a^{2}\right\rangle_{0}\left\langle b\right\rangle_{0}- 2\left\langle a\right\rangle_{0}\left\langle ab\right\rangle_{0}\nonumber\\
   && +2\left\langle a\right\rangle_{0}^{2}\left\langle b\right\rangle_{0}.\nonumber\label{e:4.9d}
\end{eqnarray}
The term $C\left(a,b\right)^{1,1}_{0}$ can also be described as the \textit{covariance} of $a$ with $b$. 

%%%%%%%%%%%%%%%%%%%%%%%%%%%%%%%%%%%%%%%%%%%%%%%%%%%%%%%%%%%%%%%%%%%%%%%%%%%%%%%%%%%%%%%
%%%%%%%%%%%%%%%%%%%%%%%%%%%%%%%%%%%%%%%%%%%%%%%%%%%%%%%%%%%%%%%%%%%%%%%%%%%%%%%%%%%%%%%
%%%%%%%%%%%%%%%%%%%%%%%%%%%%%%%%%%%%%%%%%%%%%%%%%%%%%%%%%%%%%%%%%%%%%%%%%%%%%%%%%%%%%%%
\subsection{First order moment}\label{s:2.1}
The first central moment of the TOF is
\begin{equation}
    \langle\tau\rangle_{0} = \langle t\rangle_{0}+x\left\langle v^{-1}\right\rangle_{0},\label{e:2.2}
\end{equation}
The right-hand side shows that this moment is a linear function of distance $x$. Given measurements of the first central moment on two collinear detectors at different distances, we can calculate the mean time of neutron production, $\langle t\rangle_{0}$, (a quantity that is frequently referred to as "bang-time" in ICF) and the mean of the inverse neutron velocity, $\left\langle v^{-1}\right\rangle_{0}$. This latter quantity will be examined in more detail in section \ref{s:3}. For now, we note that the two quantities are either independent of neutron velocity effects, in the case of $\langle t\rangle_{0}$, or independent of neutron production time effects, in the case of $\left\langle {v}^{-1}\right\rangle_{0}$. These quantities are what would be inferred from the first moment of the TOF on a single detector located at $x = 0$ and $x = \infty$, respectively. However, \eqref{e:2.2} shows that they can be measured by placing two collinear detectors at arbitrary distances from the neutron source.

%%%%%%%%%%%%%%%%%%%%%%%%%%%%%%%%%%%%%%%%%%%%%%%%%%%%%%%%%%%%%%%%%%%%%%%%%%%%%%%%%%%%%%%
%%%%%%%%%%%%%%%%%%%%%%%%%%%%%%%%%%%%%%%%%%%%%%%%%%%%%%%%%%%%%%%%%%%%%%%%%%%%%%%%%%%%%%%
%%%%%%%%%%%%%%%%%%%%%%%%%%%%%%%%%%%%%%%%%%%%%%%%%%%%%%%%%%%%%%%%%%%%%%%%%%%%%%%%%%%%%%%

\subsection{Second order moment}\label{s:2.2}
The second central moment of the TOF is
\begin{equation}
C\left(\tau\right)^{2}_{0}=C\left(t,v^{-1}\right)^{2,0}_{0}+2xC\left(t,v^{-1}\right)^{1,1}_{0}+x^{2}C\left(t,v^{-1}\right)^{0,2}_{0}.\label{e:2.3}
\end{equation}
The right-hand side shows that this moment is a quadratic function of detector distance $x$. Given measurements of the second central moment of TOF on three collinear detectors, each at a different distance from the neutron source , we can calculate the three inverse-velocity co-moments shown in \eqref{e:2.3}. The first of these quantities, $C\left(t,v^{-1}\right)^{2,0}_{0}$, is the variance of the neutron production time (a quantity that is closely related to the "burn-width" which is measured in ICF experiments). The second inverse-velocity co-moment is the covariance of inverse neutron velocity with neutron production time and the third is the variance of inverse neutron velocity. 

%%%%%%%%%%%%%%%%%%%%%%%%%%%%%%%%%%%%%%%%%%%%%%%%%%%%%%%%%%%%%%%%%%%%%%%%%%%%%%%%%%%%%%%
%%%%%%%%%%%%%%%%%%%%%%%%%%%%%%%%%%%%%%%%%%%%%%%%%%%%%%%%%%%%%%%%%%%%%%%%%%%%%%%%%%%%%%%
%%%%%%%%%%%%%%%%%%%%%%%%%%%%%%%%%%%%%%%%%%%%%%%%%%%%%%%%%%%%%%%%%%%%%%%%%%%%%%%%%%%%%%%

\subsection{Third order moment}\label{s:2.3}
The third central moment of the TOF is
\begin{eqnarray}
    C\left(\tau\right)^{3}_{0} &=& C\left(t,v^{-1}\right)^{3,0}_{0}+3xC\left(t,v^{-1}\right)^{2,1}_{0}\nonumber\\
&&+3x^{2}C\left(t,v^{-1}\right)^{1,2}_{0}+x^{3}C\left(t,v^{-1}\right)^{0,3}_{0}.\label{e:2.4}
\end{eqnarray}
This moment is a cubic function of distance $x$. Given measurements of the third central moment of TOF on four collinear detectors, each at a different distance from the neutron source, we can calculate the four inverse-velocity co-moments of \eqref{e:2.4}. Of particular interest is the third of these co-moments, $C\left(t,v^{-1}\right)^{1,2}_{0}$, which represents a covariance of neutron production time with inverse neutron velocity squared. As will be shown in section \ref{s:4.3}, this particular co-moment contains information on how the ion temperature changes with time during neutron production.

%%%%%%%%%%%%%%%%%%%%%%%%%%%%%%%%%%%%%%%%%%%%%%%%%%%%%%%%%%%%%%%%%%%%%%%%%%%%%%%%%%%%%%%
%%%%%%%%%%%%%%%%%%%%%%%%%%%%%%%%%%%%%%%%%%%%%%%%%%%%%%%%%%%%%%%%%%%%%%%%%%%%%%%%%%%%%%%
%%%%%%%%%%%%%%%%%%%%%%%%%%%%%%%%%%%%%%%%%%%%%%%%%%%%%%%%%%%%%%%%%%%%%%%%%%%%%%%%%%%%%%%
%%%%%%%%%%%%%%%%%%%%%%%%%%%%%%%%%%%%%%%%%%%%%%%%%%%%%%%%%%%%%%%%%%%%%%%%%%%%%%%%%%%%%%%
%%%%%%%%%%%%%%%%%%%%%%%%%%%%%%%%%%%%%%%%%%%%%%%%%%%%%%%%%%%%%%%%%%%%%%%%%%%%%%%%%%%%%%%

\section{The inverse-velocity co-moments}\label{s:3}
The set of inverse-velocity co-moments is an infinite set with elements $\langle t\rangle_{0}$, $\langle v^{-1}\rangle_{0}$, and $C\left(t,v^{-1}\right)^{j,k}_{0}$, where $j,k\in\mathbb{N}_{0}$, $j+k>1$. The order of an inverse-velocity co-moment is $j+k$, with $\langle t\rangle_{0}$ and $\langle v^{-1}\rangle_{0}$ being first order. In this section we carry out an analysis of the inverse-velocity co-moments and their relationship with ion distribution functions with the goal of gaining insights into the physical information contained within them. 

The analysis focuses on two specific features of the inverse-velocity co-moment expressions: the dependence on powers of inverse neutron velocity and the presence of a normalization factor, $\left\langle 1\right\rangle$, which depends on the ion distribution functions. We use expansions in small parameters, detailed in sections \ref{s:3.1} and \ref{s:3.2} below, respectively, to make these terms more amenable to interpretation. Before embarking on these expansions, we note that we can use the binomial theorem and linearity of $\left\langle\right\rangle_{0}$ to write \eqref{e:4.9a} 
as
\begin{eqnarray}
C\left(t,v^{-1}\right)^{j,k}_{0}&=&\sum_{p=0}^{j}\sum_{q=0}^{k}\binom{j}{p}\binom{k}{q}\left(-1\right)^{j+k-p-q}\nonumber\\
&&\times\left\langle t\right\rangle_{0}^{j-p}\left\langle v^{-1}\right\rangle_{0}^{k-q}\left\langle t^{p}v^{-q}\right\rangle_{0}.\label{e:3.1a}
\end{eqnarray}
From this we can see that interpretation of the inverse-velocity co-moments will require understanding the terms $\left\langle t\right\rangle_{0}^{j-p}$, $\left\langle v^{-1}\right\rangle_{0}^{k-q}$, and $\left\langle t^{p}v^{-q}\right\rangle_{0}$, a task that will be achieved by means of the expansions.

%%%%%%%%%%%%%%%%%%%%%%%%%%%%%%%%%%%%%%%%%%%%%%%%%%%%%%%%%%%%%%%%%%%%%%%%%%%%%%%%%%%%%%%
%%%%%%%%%%%%%%%%%%%%%%%%%%%%%%%%%%%%%%%%%%%%%%%%%%%%%%%%%%%%%%%%%%%%%%%%%%%%%%%%%%%%%%%
%%%%%%%%%%%%%%%%%%%%%%%%%%%%%%%%%%%%%%%%%%%%%%%%%%%%%%%%%%%%%%%%%%%%%%%%%%%%%%%%%%%%%%%

\subsection{Expansions of velocity terms}\label{s:3.1}
Equation \eqref{e:1.4.2a} is an expression for the velocity of a neutron emitted towards a detector as a function of the CM velocity $\vec{v_{c}}$ and relative velocity magnitude $v_{r}$ of the reacting ion pair. We can write \eqref{e:1.4.2a} as
\begin{eqnarray}
v &=& v_{0}\left(\beta+\sqrt{1+\alpha-\zeta+\beta^{2}}\right),\label{e:3.2a}\\
\alpha&=& \nu\frac{ v_{r}^{2}}{v_{0}^{2}},\quad \beta = \frac{\vec{v_{c}}\cdot\hat{v}}{v_{0}},\quad \zeta = \frac{v_{c}^{2}}{v_{0}^{2}},\label{e:3.2b}
\end{eqnarray}
where $v_{0}$ is a constant representing the  neutron velocity in the CM frame when $v_{r} = 0$. Appendix \ref{s:A1.1} gives values of the constant. 

We now make an important, but not particularly restrictive, assumption about the ion distribution functions. We assume that all reacting ion pairs in the distribution functions have $v_{r}<<v_{0}$ and $v_{c}<<v_{0}$, from which we conclude
\begin{equation}
\alpha,\,\beta,\,\zeta << 1.\label{e:3.3}
\end{equation}
Our assumption is rather easily justified for ICF conditions in which ion temperatures are $\sim 10\,keV$ and neutron energies are at least $\sim 1\,MeV$.

Now, given the assumed smallness of $\alpha$, $\beta$ and $\zeta$, any function of neutron velocity $v$ can be expanded in terms of these small parameters. One such expansion that is particular interest to us is given by \eqref{e:3.4} below. The function on the left-hand side of \eqref{e:3.4} that we have expanded is the Jacobian determinant for the differential cross section, discussed in appendix \ref{s:A1.2}, multiplied by a factor of $v^{-q}$. From the definition of $R_{k}$ in \eqref{e:1.4.2b}, we can see that this function features in the calculation of moments of the form $\left\langle v^{-q}\right\rangle$. The purpose of the expansion is to transform from a function that is nonlinearly dependent on $\alpha$, $\beta$ and $\zeta$ to a function that is a summation of powers of these variables. Given \eqref{e:3.3}, this summation will be a converging series.

\begin{widetext}
\begin{eqnarray}
    \frac{1}{v^{q}}\frac{v^{2}}{u\sqrt{u^{2}-v_{c}^{2}+\left(\vec{v_{c}}\cdot\hat{v}\right)^{2}}} &\approx& v_{0}^{-q}\left[1+\left(2-q\right)\beta+\frac{\left(q-1\right)\left(q-3\right)}{2}\beta^{2}+\frac{\left(q-1\right)}{2}\zeta-\frac{q}{2}\alpha+\mathcal{O}\left(v_{0}^{-3}\right)\right],\quad q\in\mathbb{N}_{0},\label{e:3.4}\\
    \left\langle g\right\rangle_{\sim} &=& \int g v_{r}\frac{d\sigma}{d\Omega_{cm}}F\left(t,\vec{v_{c}},\vec{v_{r}}\right)d^{3}\vec{v_{c}}d^{3}\vec{v_{r}}dVdt,\label{e:3.5a}\\
    \left\langle g\right\rangle_{\emptyset} &=& \frac{\left\langle g\right\rangle_{\sim}}{\left\langle 1\right\rangle_{\sim}}, \label{e:3.5b}\\
    \left\langle t^{p}v^{-q}\right\rangle &\approx& \frac{\left\langle 1\right\rangle_{\sim}}{v_{0}^{q}}\left[\left\langle t^{p}\right\rangle_{\emptyset}+\left(2-q\right)\left\langle\beta t^{p}\right\rangle_{\emptyset}+\frac{\left(q-1\right)\left(q-3\right)}{2}\left\langle\beta^{2}t^{p}\right\rangle_{\emptyset}\right.\nonumber\\
    &&\left.\qquad\qquad+\frac{\left(q-1\right)}{2}\left\langle\zeta t^{p}\right\rangle_{\emptyset}-\frac{q}{2}\left\langle\alpha t^{p}\right\rangle_{\emptyset}+\mathcal{O}\left(v_{0}^{-3}\right)\right],\qquad p,q\in\mathbb{N}_{0}.\label{e:3.6}
\end{eqnarray}
\end{widetext}
In order to exploit our expansion, we must first define the quantities \eqref{e:3.5a} and \eqref{e:3.5b} above. These new quantities should be contrasted with \eqref{e:1.4.3} and \eqref{e:1.4.4}, respectively. A physical interpretation will be offered in section \ref{s:5}, but for now we may simply consider \eqref{e:3.5a} and \eqref{e:3.5b} to be a set of co-moments that are compatible with \eqref{e:3.4}.

Now, with the expansion \eqref{e:3.4}, the definitions \eqref{e:1.4.3}, \eqref{e:3.5a}-\eqref{e:3.5b}, and the linearity property \eqref{en:1}, we obtain \eqref{e:3.6} above. Since \eqref{e:3.4} is a converging series it is permissible to carry out term by term integration in order to obtain \eqref{e:3.6}. The inclusion of the $t^{p}$ factor in \eqref{e:3.6} has been rather trivial since it is unaffected by the expansion \eqref{e:3.4}. Some cases of particular interest for \eqref{e:3.6} include: $p=0$, $q=0$ corresponding to $\left\langle 1\right\rangle$ (we note that, by definition, $\left\langle 1\right\rangle_{0}=\left\langle 1\right\rangle_{\emptyset}=1$), $p=1$, $q=0$ corresponding to $\left\langle t\right\rangle$, and $p=0$, $q=1$ corresponding to $\left\langle v^{-1}\right\rangle$. 

%%%%%%%%%%%%%%%%%%%%%%%%%%%%%%%%%%%%%%%%%%%%%%%%%%%%%%%%%%%%%%%%%%%%%%%%%%%%%%%%%%%%%%%
%%%%%%%%%%%%%%%%%%%%%%%%%%%%%%%%%%%%%%%%%%%%%%%%%%%%%%%%%%%%%%%%%%%%%%%%%%%%%%%%%%%%%%%
%%%%%%%%%%%%%%%%%%%%%%%%%%%%%%%%%%%%%%%%%%%%%%%%%%%%%%%%%%%%%%%%%%%%%%%%%%%%%%%%%%%%%%%

\subsection{Expansion of the normalization factor}\label{s:3.2}
The definition of $\left\langle g\right\rangle_{0}$ in \eqref{e:1.3.2} shows that $\left\langle 1\right\rangle^{-1}$ is required as a normalization factor. Given the assumption \eqref{e:3.3} and the definition \eqref{e:3.5b}, we see that terms such as $\left\langle\beta^{n}\right\rangle_{\emptyset}$, $\left\langle\alpha^{n}\right\rangle_{\emptyset}$,$\left\langle\zeta^{n}\right\rangle_{\emptyset}$,$\ldots$ are all $<< 1$, for $n\in\mathbb{N}$. Therefore, we can approximate the normalization factor by first using \eqref{e:3.6} with $p=0$, $q=0$ to approximate $\left\langle 1\right\rangle^{-1}$ and then expanding the result in this new infinite series of small parameters. This gives
\begin{comment}
\begin{equation}
    \frac{1}{\langle 1\rangle} \approx \frac{1}{\langle 1\rangle_{\sim}}\left[1-2\left\langle\beta\right\rangle_{\emptyset}+\frac{5}{2}\left\langle\beta^{2}\right\rangle_{\emptyset}+\frac{1}{2}\left\langle\zeta\right\rangle_{\emptyset}+\mathcal{O}\left(v_{0}^{-3}\right)\right].\label{e:3.7}
\end{equation}
\end{comment}
\begin{equation}
    \frac{1}{\langle 1\rangle} \approx \frac{1}{\langle 1\rangle_{\sim}}\left[1-2\left\langle\beta\right\rangle_{\emptyset}-\frac{3}{2}\left\langle\beta^{2}\right\rangle_{\emptyset}+4\left\langle\beta\right\rangle^{2}_{\emptyset}+\frac{1}{2}\left\langle\zeta\right\rangle_{\emptyset}+\mathcal{O}\left(v_{0}^{-3}\right)\right].\label{e:3.7}
\end{equation}
The significance of \eqref{e:3.7} is that our normalization factor has now been approximated as a converging series rather than being a denominator, a form which is much more amenable to analysis.

%%%%%%%%%%%%%%%%%%%%%%%%%%%%%%%%%%%%%%%%%%%%%%%%%%%%%%%%%%%%%%%%%%%%%%%%%%%%%%%%%%%%%%%
%%%%%%%%%%%%%%%%%%%%%%%%%%%%%%%%%%%%%%%%%%%%%%%%%%%%%%%%%%%%%%%%%%%%%%%%%%%%%%%%%%%%%%%
%%%%%%%%%%%%%%%%%%%%%%%%%%%%%%%%%%%%%%%%%%%%%%%%%%%%%%%%%%%%%%%%%%%%%%%%%%%%%%%%%%%%%%%

\subsection{Expressions for inverse-velocity co-moments}\label{s:3.3}
We have completed the necessary expansions for interpreting the inverse-velocity co-moments. We now assemble our results to come up with expressions for the inverse-velocity co-moments in the form of a converging infinite series. First, we can multiply \eqref{e:3.6} by \eqref{e:3.7} to get \eqref{e:3.8} below, an approximation for $\left\langle t^{p}v^{-q}\right\rangle_{0}$. Next we can use the multinomial theorem (see appendix \ref{s:A5.1}) to calculate powers of \eqref{e:3.8}, which are then ordered as a converging series. The results of this for the cases $p=1,\,q=0$ and $p=0,\,q=1$ are shown in \eqref{e:3.9a} and \eqref{e:3.9b}
\begin{widetext}

\begin{eqnarray}
      \left\langle t^{p}v^{-q}\right\rangle_{0} &\approx& v_{0}^{-q}\left[\left\langle t^{p}\right\rangle_{\emptyset}+\left(2-q\right)\left\langle\beta t^{p}\right\rangle_{\emptyset}-2\left\langle\beta \right\rangle_{\emptyset}\left\langle  t^{p}\right\rangle_{\emptyset}\right.+\frac{\left(q-1\right)\left(q-3\right)}{2}\left\langle\beta^{2}t^{p}\right\rangle_{\emptyset}-\frac{3}{2}\left\langle\beta^{2}\right\rangle_{\emptyset}\left\langle t^{p}\right\rangle_{\emptyset}-2\left(2-q\right)\left\langle\beta\right\rangle_{\emptyset}\left\langle\beta t^{p}\right\rangle_{\emptyset}\nonumber\\
  &&\left.+4\left\langle\beta \right\rangle_{\emptyset}^{2}\left\langle t^{p}\right\rangle_{\emptyset}-\frac{q}{2}\left\langle\alpha t^{p}\right\rangle_{\emptyset}+\frac{\left(q-1\right)}{2}\left\langle\zeta t^{p}\right\rangle_{\emptyset}+\frac{1}{2}\left\langle\zeta \right\rangle_{\emptyset}\left\langle  t^{p}\right\rangle_{\emptyset}+\mathcal{O}\left(v_{0}^{-3}\right)\right],\label{e:3.8}\\
   \left\langle t\right\rangle_{0}^{n} &\approx& \left\langle t\right\rangle_{\emptyset}^{n}+2n\left\langle t\right\rangle_{\emptyset}^{n-1}C\left(t,\beta\right)_{\emptyset}^{1,1}+\frac{3}{2}n\left\langle t\right\rangle_{\emptyset}^{n-1}C\left(t,\beta^{2}\right)_{\emptyset}^{1,1}-4n\left\langle t\right\rangle_{\emptyset}^{n-1}\left\langle\beta\right\rangle_{\emptyset}C\left(t,\beta\right)_{\emptyset}^{1,1}\nonumber\\
  &&+2n\left(n-1\right)\left\langle t\right\rangle_{\emptyset}^{n-2}\left(C\left(t,\beta\right)_{\emptyset}^{1,1}\right)^{2}-\frac{1}{2}n\left\langle t\right\rangle_{\emptyset}^{n-1}C\left(t,\zeta\right)_{\emptyset}^{1,1}+\mathcal{O}\left(v_{0}^{-3}\right),\label{e:3.9a}\\
\left\langle v^{-1}\right\rangle_{0}^{n} &\approx& v_{0}^{-n}\left[1-n\left\langle\beta \right\rangle_{\emptyset}-\frac{3}{2}n\left\langle\beta^{2}\right\rangle_{\emptyset}+\frac{n\left(n+3\right)}{2}\left\langle\beta \right\rangle_{\emptyset}^{2}-\frac{n}{2}\left\langle\alpha \right\rangle_{\emptyset}+\frac{n}{2}\left\langle\zeta \right\rangle_{\emptyset}+\mathcal{O}\left(v_{0}^{-3}\right)\right],\label{e:3.9b}
\end{eqnarray}

\end{widetext}

Now, combining \eqref{e:3.1a}, \eqref{e:3.8},\eqref{e:3.9a}, and \eqref{e:3.9b} gives us an approximation for the inverse-velocity co-moments in the form of a converging series as given by \eqref{e:3.10} below. This expression represents one of the key results of our work. It is the inverse-velocity co-moments in the form of a series of increasing powers of the terms $\alpha$, $\beta$ and $\zeta$. Given the smallness of these terms, the importance of these terms decreases as the powers increase, and so an inverse-velocity co-moment will be most sensitive to the leading order, non-zero terms in this expansion. In \eqref{e:3.10}, we have chosen to expand up to $\mathcal{O}\left(v_{0}^{-2}\right)$, but this could be increased as desired.

\begin{widetext}

\begin{eqnarray}
v_{0}^{k}C\left(t,v^{-1}\right)^{j,k}_{0}&\approx&\sum_{p=0}^{j}\sum_{q=0}^{k}\binom{j}{p}\binom{k}{q}\left(-1\right)^{j+k-p-q}\left[\left\langle t^{p}\right\rangle_{\emptyset}\left\langle  t\right\rangle_{\emptyset}^{j-p}\right.\nonumber\\
&& +\left(2-q\right)\left\langle  t\right\rangle_{\emptyset}^{j-p}C\left(t^{p},\beta\right)_{\emptyset}^{1,1}+2\left(j-p\right)\left\langle  t^{p}\right\rangle_{\emptyset}\left\langle  t\right\rangle_{\emptyset}^{j-p-1}C\left(t,\beta\right)_{\emptyset}^{1,1}-k\left\langle  \beta \right\rangle_{\emptyset}\left\langle  t^{p}\right\rangle_{\emptyset}\left\langle  t\right\rangle_{\emptyset}^{j-p}\nonumber\\
&&+\frac{3}{2}\left(j-p\right)\left\langle  t^{p}\right\rangle_{\emptyset}\left\langle  t\right\rangle_{\emptyset}^{j-p-1}C\left(t,\beta^{2}\right)_{\emptyset}^{1,1}\nonumber\\
&&+\frac{1}{2}\left\langle  t\right\rangle_{\emptyset}^{j-p}\left(\left(q-1\right)\left(q-3\right)\left\langle\beta^{2}t^{p}\right\rangle_{\emptyset}-3\left(k-q+1\right)\left\langle\beta^{2}\right\rangle_{\emptyset}\left\langle t^{p}  \right\rangle_{\emptyset}\right)\nonumber\\
&&+\frac{\left(q-1\right)}{2}\left\langle t\right\rangle_{\emptyset}^{j-p}C\left(t^{p},\zeta\right)_{\emptyset}^{1,1}-\frac{1}{2}\left(j-p\right)\left\langle t^{p}\right\rangle_{\emptyset}\left\langle t\right\rangle_{\emptyset}^{j-p-1}C\left(t,\zeta\right)_{\emptyset}^{1,1}+\frac{k}{2}\left\langle\zeta\right\rangle_{\emptyset}\left\langle t^{p}\right\rangle_{\emptyset}\left\langle t\right\rangle_{\emptyset}^{j-p}  \nonumber\\
&&-\frac{q}{2}\left\langle t\right\rangle_{\emptyset}^{j-p}C\left(t^{p},\alpha\right)_{\emptyset}^{1,1}-\frac{k}{2}\left\langle\alpha\right\rangle_{\emptyset}\left\langle t^{p}\right\rangle_{\emptyset}\left\langle t\right\rangle_{\emptyset}^{j-p} \nonumber\\
&&+2\left(j-p\right)\left\langle t^{p}\right\rangle_{\emptyset}\left\langle t\right\rangle_{\emptyset}^{j-p-2}C\left(t,\beta\right)_{\emptyset}^{1,1}\left(\left(j-p-1\right)C\left(t,\beta\right)_{\emptyset}^{1,1}-2\left\langle t\right\rangle_{\emptyset}\left\langle \beta\right\rangle_{\emptyset}\right)\nonumber\\
&&+2\left(j-p\right)\left\langle t\right\rangle_{\emptyset}^{j-p-1}C\left(t,\beta\right)_{\emptyset}^{1,1}\left(\left(2-q\right)C\left(t^{p},\beta\right)_{\emptyset}^{1,1}-k\left\langle t^{p}\right\rangle_{\emptyset}\left\langle \beta\right\rangle_{\emptyset}\right)\nonumber\\
&&+\left\langle t\right\rangle_{\emptyset}^{j-p}\left\langle \beta\right\rangle_{\emptyset}\left(\left(2q-4+\left(k-q\right)\left(q-2\right)\right)C\left(t^{p},\beta\right)_{\emptyset}^{1,1}+\frac{1}{2}\left(4q+\left(k-q\right)\left(k+q+3\right)\right)\left\langle t^{p}\right\rangle_{\emptyset}\left\langle \beta\right\rangle_{\emptyset}\right)\nonumber\\
&&\left.+\mathcal{O}\left(v_{0}^{-3}\right)\right].\label{e:3.10}
\end{eqnarray}
\end{widetext}

%%%%%%%%%%%%%%%%%%%%%%%%%%%%%%%%%%%%%%%%%%%%%%%%%%%%%%%%%%%%%%%%%%%%%%%%%%%%%%%%%%%%%%%
%%%%%%%%%%%%%%%%%%%%%%%%%%%%%%%%%%%%%%%%%%%%%%%%%%%%%%%%%%%%%%%%%%%%%%%%%%%%%%%%%%%%%%%
%%%%%%%%%%%%%%%%%%%%%%%%%%%%%%%%%%%%%%%%%%%%%%%%%%%%%%%%%%%%%%%%%%%%%%%%%%%%%%%%%%%%%%%

\subsection{Examples of inverse-velocity co-moments}\label{s:3.4}
In this subsection, we offer some analyses and comments on specific inverse-velocity co-moments, and consider the behaviour of \eqref{e:3.10} for a range of $j$ and $k$ values.

\subsubsection{First order inverse-velocity co-moments}
Approximations for the first order co-moments $\left\langle t\right\rangle_{0}$ and $\left\langle v^{-1}\right\rangle_{0}$ can be obtained from \eqref{e:3.8} by setting $p=1$, $q=0$ and $p=0$, $q=1$, respectively. This is equivalent to setting $n=1$ in \eqref{e:3.9a} and \eqref{e:3.9b}, respectively. As discussed in section \ref{s:2.1}, $\left\langle t\right\rangle_{0}$ is the mean time of neutron production for neutrons emitted in the detector direction. From \eqref{e:3.9a} we see that this is approximately equal to $\left\langle t\right\rangle_{\emptyset}$ with a second order correction arising from the covariance of $\beta$ with production time $t$. According to \eqref{e:3.2b}, $\beta$ is the normalized component of the CM velocity vector directed towards the detectors. A positive value for this covariance indicates that, on average, $\vec{v_{c}}\cdot\hat{v}$ of reacting pairs is increasing with time and so $\left\langle t\right\rangle_{0}$ will have a larger value. However, since $\beta$ depends on the direction of neutron emission, there will be other directions for which the covariance will be negative giving a reduced $\left\langle t\right\rangle_{0}$ in that direction. This is particularly clear if we consider antipodal directions of neutron emission, which we can denote by $\hat{v}$ and $-\hat{v}$. The covariances of $\beta$ with $t$ will always be of equal magnitudes and opposite signs in these two directions. We return to the antipodal detector scenario in section \ref{s:7}.    

Meanwhile, $\left\langle v^{-1}\right\rangle_{0}$ is the mean of the inverse neutron velocity, which is not necessarily equal to the inverse of the mean neutron velocity, that is $\left\langle v^{-1}\right\rangle_{0} \neq \left\langle v\right\rangle_{0}^{-1}$. From \eqref{e:3.9b}, we see that $\left\langle v^{-1}\right\rangle_{0}$ is approximately $v_{0}^{-1}$ with a first order correction of $-\left\langle \beta\right\rangle_{\emptyset}$. The presence of the minus sign here is because we are considering the inverse neutron velocity, a quantity which is reduced if the value of $\vec{v_{c}}\cdot\hat{v}$ is positive.

\subsubsection{Time-integrated co-moments ($j = 0$, $k>1$)}
We next consider the scenario with $j = 0$, $k>1$. We refer to this subset of the inverse-velocity co-moments as \textit{time-integrated}, since the co-moments are independent of the neutron production time variable $t$. Using the binomial identity from appendix \ref{s:A5.2}, we can see that any term in the summation of \eqref{e:3.10} which does not contain a factor $q$ will be $0$. Therefore, the time-integrated co-moments are 
\begin{comment}
\begin{eqnarray}
&&v_{0}^{k}C\left(t,v^{-1}\right)^{0,k}_{0}\approx\left(-1\right)^{k+1}\left[4C\left(t,\beta\right)_{\emptyset}^{0,2}+\mathcal{O}\left(v_{0}^{-3}\right)\right]\nonumber\\
&&+\sum_{q=0}^{k}\binom{k}{q}\left(-1\right)^{k-q}\left[\frac{q}{2}\left(q-9\right)C\left(t,\beta\right)^{0,2}_{\emptyset}+\mathcal{O}\left(v_{0}^{-3}\right)\right].\label{e:3.17}
\end{eqnarray}
\end{comment}
\begin{eqnarray}
v_{0}^{k}C\left(t,v^{-1}\right)^{0,k}_{0}&&\approx\sum_{q=0}^{k}\binom{k}{q}\left(-1\right)^{k-q}\nonumber\\
&&\times\left[\frac{q}{2}\left(q-1\right)C\left(t,\beta\right)^{0,2}_{\emptyset}+\mathcal{O}\left(v_{0}^{-3}\right)\right].\label{e:3.17}
\end{eqnarray}
Further applications of the binomial identity show that for $k>2$, the leading order terms in the summation of \eqref{e:3.10} will be $\mathcal{O}\left(v_{0}^{-3}\right)$ or greater.

As well as the time-integrated co-moments, \textit{velocity-integrated} inverse-velocity co-moments also exist. These are the co-moments for which $j > 1$, $k = 0$, and are of the form $C\left(t,v^{-1}\right)^{j,0}_{0}$. In contrast to the time-integrated co-moments, for the velocity-integrated co-moments many of the terms on the right-hand side of \eqref{e:3.10} will remain non-zero, due to the presence of a $p$ exponent in terms such as $\left\langle t^{p}\right\rangle_{\emptyset}$ and $\left\langle  t\right\rangle_{\emptyset}^{j-p}$. An exception is terms containing $\alpha$ in \eqref{e:3.10} which are both eliminated when $k = 0$.

\subsubsection{Co-moments with $k\geq 1$}
The binomial identity used to obtain \eqref{e:3.17} can also be applied to a larger subset of co-moments, namely, those for which $j\geq 1$, $k\geq 1$. The identity means terms in \eqref{e:3.10} that do not include a factor of $q^{k}$, or an exponent larger than $k$, will be $0$. This includes the leading order term in \eqref{e:3.10}, 
$\left\langle t^{p}\right\rangle_{\emptyset}\left\langle  t\right\rangle_{\emptyset}^{j-p}$, and if $k>1$, terms of $\mathcal{O}\left(v_{0}^{-1}\right)$ will also be $0$. As an example, for $k=2$ we get
\begin{comment}
\begin{eqnarray}
v_{0}^{2}&&C\left(t,v^{-1}\right)^{j,2}_{0}\approx\left(-1\right)^{k+1}\left[4C\left(t,\beta\right)_{\emptyset}^{0,2}+\mathcal{O}\left(v_{0}^{-3}\right)\right]\nonumber\\
&&+\sum_{p=0}^{j}\binom{j}{p}\left(-1\right)^{j-p}\left[
\left\langle\beta^{2}t^{p}\right\rangle_{\emptyset}\left\langle t\right\rangle_{\emptyset}^{j-p}+\left\langle\beta\right\rangle_{\emptyset}^{2}\left\langle t^{p}\right\rangle_{\emptyset}\left\langle t\right\rangle_{\emptyset}^{j-p}\right.\nonumber\\
&&\left.-2\left\langle\beta t^{p}\right\rangle_{\emptyset}\left\langle\beta\right\rangle_{\emptyset}\left\langle t\right\rangle_{\emptyset}^{j-p} +\mathcal{O}\left(v_{0}^{-3}\right)\right],\label{e:3.16}
\end{eqnarray}
\end{comment}
\begin{eqnarray}
v_{0}^{2}C\left(t,v^{-1}\right)^{j,2}_{0}&&\approx\sum_{p=0}^{j}\binom{j}{p}\left(-1\right)^{j-p}\nonumber\\
&&\times\left[
\left\langle\beta^{2}t^{p}\right\rangle_{\emptyset}\left\langle t\right\rangle_{\emptyset}^{j-p}+\left\langle\beta\right\rangle_{\emptyset}^{2}\left\langle t^{p}\right\rangle_{\emptyset}\left\langle t\right\rangle_{\emptyset}^{j-p}\right.\nonumber\\
&&\left.-2\left\langle\beta t^{p}\right\rangle_{\emptyset}\left\langle\beta\right\rangle_{\emptyset}\left\langle t\right\rangle_{\emptyset}^{j-p} +\mathcal{O}\left(v_{0}^{-3}\right)\right],\label{e:3.16}
\end{eqnarray}
from which we can see that the leading order terms are $\mathcal{O}\left(v_{0}^{-2}\right)$.

%%%%%%%%%%%%%%%%%%%%%%%%%%%%%%%%%%%%%%%%%%%%%%%%%%%%%%%%%%%%%%%%%%%%%%%%%%%%%%%%%%%%%%%
%%%%%%%%%%%%%%%%%%%%%%%%%%%%%%%%%%%%%%%%%%%%%%%%%%%%%%%%%%%%%%%%%%%%%%%%%%%%%%%%%%%%%%%
%%%%%%%%%%%%%%%%%%%%%%%%%%%%%%%%%%%%%%%%%%%%%%%%%%%%%%%%%%%%%%%%%%%%%%%%%%%%%%%%%%%%%%%
%%%%%%%%%%%%%%%%%%%%%%%%%%%%%%%%%%%%%%%%%%%%%%%%%%%%%%%%%%%%%%%%%%%%%%%%%%%%%%%%%%%%%%%
%%%%%%%%%%%%%%%%%%%%%%%%%%%%%%%%%%%%%%%%%%%%%%%%%%%%%%%%%%%%%%%%%%%%%%%%%%%%%%%%%%%%%%%

\section{Inverse-velocity co-moments for Maxwellians}\label{s:4}
The approximation for inverse-velocity co-moments given by \eqref{e:3.10} is valid for arbitrary ion velocity distribution functions. The only requirement is the relations \eqref{e:3.3}. An ion distribution function of particular interest is the Maxwellian, which we consider in this section. We assume that we have two reactant ion species which have densities, $n_{1}$ and $n_{2}$, and a single ion temperature, $T$, and bulk fluid velocity, $\vec{v_{f}}$. Given these distributions, the joint distribution of $\vec{v_{c}}$ and $\vec{v_{r}}$ is
\begin{eqnarray}
    F&&\left(t,\vec{v_{c}},\vec{v_{r}}\right) = n_{1}n_{2}\left(\frac{\sqrt{m_{1}m_{2}}}{2\pi T}\right)^{3}\exp\left(-\frac{m_{12}}{2T}v_{r}^{2}\right)\times\nonumber\\
    &&\exp\left(-\frac{\left(m_{1}+m_{2}\right)}{2T}\left(\vec{v_{c}}-\vec{v_{f}}\right)^{2}\right)d^{3}\vec{v_{c}}d^{3}\vec{v_{r}},\label{e:4.1}
\end{eqnarray}
where $m_{12}=m_{1}m_{2}/\left(m_{1}+m_{2}\right)$ is the reduced mass of pairs of particles from species $1$ and $2$. We note that the paramters $T$, $\vec{v_{f}}$, $n_{1}$ and $n_{2}$ can all vary in time and space.

Given \eqref{e:4.1}, we find that the computation of $\left\langle g\right\rangle_{\sim}$, using \eqref{e:3.5a}, involves integration over several variables which are trivial or have analytic results. In contrast, computing $\left\langle g\right\rangle$ using \eqref{e:1.4.3} is more complex due to the Jacobian determinant term. In particular, we note that since \eqref{e:4.1} depends only on the magnitude of $\vec{v_{r}}$ and is independent of its direction, we can integrate the differential cross section over the azimuthal and polar angles of $\vec{v_{r}}$ to obtain total cross section $\sigma$. Making use of these results, we define a new moment $\left\langle g\right\rangle_{m}$
\begin{equation}
\left\langle g\right\rangle_{m}=\frac{\int g \frac{n_{1}n_{2}}{\left(T\right)^{3/2}}v_{r}^{3}\sigma\exp\left(-\frac{m_{12}}{2\pi T}v_{r}^{2}\right)dv_{r}dVdt}{\int \frac{n_{1}n_{2}}{\left(T\right)^{3/2}}v_{r}^{3}\sigma\exp\left(-\frac{m_{12}}{2\pi T}v_{r}^{2}\right)dv_{r}dVdt},\label{e:4.2}
\end{equation}
where the denominator here corresponds to $\left\langle 1\right\rangle_{\sim}$ for Maxwellian distributions. The notation of \eqref{e:4.9a} can also be extended to these new moments using a change of subscript, for example $C\left(a,b\right)^{j,k}_{m}$.

Now, using \eqref{e:3.5a} and \eqref{e:3.5b}, we can equate certain moments of the type $\left\langle\right\rangle_{\emptyset}$ to moments of the type $\left\langle\right\rangle_{m}$. The following results are or particular interest
\begin{eqnarray}
    \left\langle t^{p}\right\rangle_{\emptyset} &=& \left\langle t^{p}\right\rangle_{m},\label{e:4.3a}\\
    \left\langle \beta t^{p}\right\rangle_{\emptyset} &=& v_{0}^{-1}\left\langle \vec{v_{f}}\cdot\hat{v}t^{p}\right\rangle_{m},\label{e:4.3b}\\
    \left\langle \beta^{2} t^{p}\right\rangle_{\emptyset} &=& v_{0}^{-2}\left\langle \left(\frac{T}{m_{1}+m_{2}}+\left(\vec{v_{f}}\cdot\hat{v}\right)^{2}\right)t^{p}\right\rangle_{m},\label{e:4.3c}\\
    \left\langle \zeta t^{p}\right\rangle_{\emptyset} &=& v_{0}^{-2}\left\langle \left(\frac{3T}{m_{1}+m_{2}}+v_{f}^{2}\right)t^{p}\right\rangle_{m},\label{e:4.3d}\\
    \left\langle \alpha t^{p}\right\rangle_{\emptyset} &=& v_{0}^{-2}\left\langle \nu v_{r}^{2}t^{p}\right\rangle_{m},\label{e:4.3e}
\end{eqnarray}
where $\nu$ is a function of particle masses defined in \eqref{en:3}. 

With relations \eqref{e:4.3a}-\eqref{e:4.3e}, we can rewrite the inverse-velocity co-moment approximation \eqref{e:3.10} in terms of $\left\langle\right\rangle_{m}$ when the ion distribution functions are Maxwellian. This will then show how the inverse-velocity co-moments are affected by ion temperature and bulk fluid velocity, and the variation of these quantities with time. In the following subsections we consider the inverse-velocity co-moments which arise in the first three central moments of the TOF, described in subsections \ref{s:2.1} to \ref{s:2.3}. In each case, we write out expressions for the inverse-velocity co-moments for Maxwellian distributions, and discuss some physical interpretations of the terms that occur in these expressions.

\subsection{First order moment for Maxwellians}\label{s:4.1}
The first central moment of the TOF is defined by \eqref{e:2.2}. It is a function of the inverse-velocity co-moments $\left\langle t\right\rangle_{0}$ and $\left\langle v^{-1}\right\rangle_{0}$. Using \eqref{e:3.8} and \eqref{e:4.3a}-\eqref{e:4.3e}, we obtain the following approximations for these co-moments for the case of Maxwellian distribution functions
\begin{widetext}
\begin{eqnarray}
\left\langle t\right\rangle_{0} &\approx& \left\langle t\right\rangle_{m}+2v_{0}^{-1}C\left(t,\vec{v_{f}}\cdot\hat{v}\right)^{1,1}_{m}+v_{0}^{-2}\left[\frac{3}{2}C\left(t,\left(\vec{v_{f}}\cdot\hat{v}\right)^{2}\right)^{1,1}_{m}-\frac{1}{2}C\left(t,v_{f}^{2}\right)^{1,1}_{m}-4\left\langle \vec{v_{f}}\cdot\hat{v}\right\rangle_{m}C\left(t,\vec{v_{f}}\cdot\hat{v}\right)^{1,1}_{m}\right]\nonumber\\
&&+\mathcal{O}\left(v_{0}^{-3}\right),\label{e:4.6a}\\
v_{0}\left\langle v^{-1}\right\rangle_{0}-1 &\approx& -v_{0}^{-1}\left\langle \vec{v_{f}}\cdot\hat{v}\right\rangle_{m}+v_{0}^{-2}\left[\frac{1}{2}\left\langle  v_{f}^{2}\right\rangle_{m}-\frac{\nu}{2}\left\langle 
 v_{r}^{2}\right\rangle_{m}-\frac{3}{2}\left\langle\left(\vec{v_{f}}\cdot\hat{v}\right)^{2}\right\rangle_{m}+2\left\langle\vec{v_{f}}\cdot\hat{v}\right\rangle_{m}^{2}\right]+\mathcal{O}\left(v_{0}^{-3}\right).\label{e:4.6b}
\end{eqnarray}
\end{widetext}
In the above equations, we use square brackets to collect together all terms of the same order with respect to $v_{0}^{-1}$. 

First, considering \eqref{e:4.6a}, we see that the leading order term here is $\left\langle t\right\rangle_{m}$, which is approximately the yield-averaged mean time of neutron production. The second order term contains the expression $C\left(t,\vec{v_{f}}\cdot\hat{v}\right)^{1,1}_{m}$. This quantity represents the covariance of the vector component of bulk fluid velocity in the detector direction with neutron production time. As shown in appendix \ref{s:A4}, it is a measure of the vector component of bulk fluid acceleration in the detector direction. If the neutron source is accelerating towards the detectors then the inferred mean neutron production time will be later.

Secondly, considering \eqref{e:4.6b}, we see that the first term on the right-hand side represents a shift in the neutron velocity spectrum due to time-integrated bulk fluid velocity. The minus sign arises from the fact that we are considering inverse neutron velocity and so a positive bulk fluid velocity component (flow directed towards the detector) will reduce the inverse-velocity co-moment. For the second order terms shown in \eqref{e:4.6b} we highlight the $\nu\left\langle v_{r}^{2}\right\rangle_{m}$ term. Measurement of this term in the neutron velocity domain was recently responsible for the identification of non-Maxwellian ion distribution functions in several experiments.\cite{Hartouni_2023, Mannion_2023,Crilly_2022} As we shall see in the upcoming subsection, the second order central moment may offer a route to inferring the time-varying behaviour of such distributions.

\subsection{Second order moment for Maxwellians}\label{s:4.2}
The second central moment of the TOF is given by \eqref{e:2.3}. It is a function of three second order inverse-velocity co-moments, for which we find the following approximations for the case of Maxwellian distributions
\begin{widetext}
\begin{eqnarray}
C\left(t,v^{-1}\right)^{2,0}_{0}&\approx&C\left(t,v^{-1}\right)^{2,0}_{m}+2v_{0}^{-1}\left[C\left(t^{2},\vec{v_{f}}\cdot\hat{v} \right)^{1,1}_{m}-2\left\langle  t\right\rangle_{m}C\left(t,\vec{v_{f}}\cdot\hat{v} \right)^{1,1}_{m}\right]+\mathcal{O}\left(v_{0}^{-2}\right)\label{e:4.7a}\\
v_{0}C\left(t,v^{-1}\right)^{1,1}_{0}&\approx&-v_{0}^{-1}C\left(t,\vec{v_{f}}\cdot\hat{v}\right)^{1,1}_{m} +v_{0}^{-2}\left[\frac{1}{2}C\left(t,v_{f}^{2}\right)^{1,1}_{m}-\frac{\nu}{2} C\left(t,v_{r}^{2}\right)^{1,1}_{m}-\frac{3}{2}C\left(t,\left(\vec{v_{f}}\cdot\hat{v}\right)^{2}\right)^{1,1}_{m}\right.\nonumber\\
&&\left.+4\left\langle\vec{v_{f}}\cdot\hat{v}\right\rangle_{m}C\left(t,\vec{v_{f}}\cdot\hat{v}\right)^{1,1}_{m} \right]+\mathcal{O}\left(v_{0}^{-3}\right).\label{e:4.7b}\\
v_{0}^{2}C\left(t,v^{-1}\right)^{0,2}_{0}&\approx&v_{0}^{-2}\left[\frac{\left\langle T\right\rangle_{m}}{\left(m_{1}+m_{2}\right)}+C\left(t,\vec{v_{f}}\cdot\hat{v}\right)^{0,2}_{m}\right]+\mathcal{O}\left(v_{0}^{-3}\right),\label{e:4.7c}
\end{eqnarray}
\end{widetext}
First, \eqref{e:4.7a} represents the variance of thermonuclear burn, as mentioned previously in section \ref{s:2.2}. Here we see that for Maxwellian ion distributions we will have second order contributions to this due to covariances of the bulk fluid velocity in the direction of neutron emission with time.

Secondly, \eqref{e:4.7b} shows that, to leading order, $C\left(t,v^{-1}\right)^{1,1}_{0}$ measures the covariance with time of the bulk fluid velocity in the direction of neutron emission. This can be used to obtain the acceleration of the neutron emitting source, as outlined in appendix \ref{s:A4}. In the second order terms of \eqref{e:4.7b} we note the presence of the $-\nu C\left(t,v_{r}^{2}\right)^{1,1}_{m}$ term. This contains information on the time-varying behaviour of certain types of non-Maxwellian ion distributions. As described in section \ref{s:7}, an antipodal configuration of two sets of three detectors could be used to measure this term by eliminating $C\left(t,\vec{v_{f}}\cdot\hat{v}\right)^{1,1}_{m}$.

Finally, \eqref{e:4.7c} provides a measure of the time-integrated ion temperature. This co-moment also includes contributions from the bulk fluid velocity variance in the direction of neutron emission, a result that is similar to that obtained from the moments method in the neutron velocity domain.

\subsection{Third order moment for Maxwellians}\label{s:4.3}
The third central moment of the TOF is given by \eqref{e:2.4}. It is a function of four third order inverse-velocity co-moments. Maxwellian approximations for these co-moments are given by \eqref{e:4.8a}-\eqref{e:4.8d} below.

First, \eqref{e:4.8a} is the third central moment of the production time, which contains second order corrections due to the bulk fluid velocity. Secondly, \eqref{e:4.8b} is, to leading order, another independent measure of the time-varying behaviour of the bulk fluid velocity.

Thirdly, and most notably, \eqref{e:4.8c} contains information on the time-varying behaviour of the ion temperature in its leading order terms. In particular, it contains the term $C\left(t,T\right)^{1,1}_{m}$ which could be used to calculate the rate of change of ion temperature with production time, as outlined in appendix \ref{s:A4}. However, it is not possible to measure this term independently from $C\left(t,\left(\vec{v_{f}}\cdot\hat{v}\right)^{2}\right)^{1,1}_{m}$. Note that the final leading order term in \eqref{e:4.8c}, $2\left\langle \vec{v_{f}}\cdot\hat{v}\right\rangle_{m}C\left(t,\vec{v_{f}}\cdot\hat{v}\right)^{1,1}_{m}$ can be estimated from the leading order terms of \eqref{e:4.6b} and \eqref{e:4.7b}.

Finally, \eqref{e:4.8d} provides information on the time-integrated ion temperature and bulk fluid velocity variance, similar to \eqref{e:4.7c}, but of a higher order.
\begin{widetext}
\begin{eqnarray}
C\left(t,v^{-1}\right)^{3,0}_{0} &\approx& C\left(t,v^{-1}\right)^{3,0}_{m}+2v_{0}^{-1}\left[C\left(t^{3},\vec{v_{f}}\cdot\hat{v}\right)^{1,1}_{m}\right.\nonumber\\
&&\left.-3\left\langle  t^{2}\right\rangle_{m}C\left(t,\vec{v_{f}}\cdot\hat{v}\right)^{1,1}_{m}-3\left\langle  t\right\rangle_{m}C\left(t,\vec{v_{f}}\cdot\hat{v}\right)^{2,1}_{m}\right]+\mathcal{O}\left(v_{0}^{-2}\right),\label{e:4.8a}\\
v_{0}^{1}C\left(t,v^{-1}\right)^{2,1}_{0}&\approx& -v_{0}^{-1}C\left(t,\vec{v_{f}}\cdot\hat{v}\right)^{2,1}_{m}+v_{0}^{-2}\left[\frac{1}{2}C\left(t,v_{f}^{2}\right)^{2,1}_{m}-\frac{\nu}{2}C\left(t,v_{r}^{2}\right)^{2,1}_{m}-\frac{3}{2}C\left(t,\left(\vec{v_{f}}\cdot\hat{v}\right)^{2}\right)^{2,1}_{m}\right.\nonumber\\
&&\left.+4\left\langle \vec{v_{f}}\cdot\hat{v}\right\rangle_{m}C\left(t,\vec{v_{f}}\cdot\hat{v}\right)^{2,1}_{m}+4C\left(t,\vec{v_{f}}\cdot\hat{v}\right)^{1,1}_{m}C\left(t,\vec{v_{f}}\cdot\hat{v}\right)^{1,1}_{m}\right]+\mathcal{O}\left(v_{0}^{-3}\right),\label{e:4.8b}\\
v_{0}^{2}C\left(t,v^{-1}\right)^{1,2}_{0}&\approx&v_{0}^{-2}\left[\frac{C\left(t,T\right)^{1,1}_{m}}{\left(m_{1}+m_{2}\right)}+C\left(t,\vec{v_{f}}\cdot\hat{v}\right)^{1,2}_{m}\right]+\mathcal{O}\left(v_{0}^{-3}\right),\label{e:4.8c}\\
v_{0}^{3}C\left(t,v^{-1}\right)^{0,3}_{0}&\approx& \mathcal{O}\left(v_{0}^{-3}\right).\label{e:4.8d}
\end{eqnarray}
\end{widetext}

%%%%%%%%%%%%%%%%%%%%%%%%%%%%%%%%%%%%%%%%%%%%%%%%%%%%%%%%%%%%%%%%%%%%%%%%%%%%%%%%%%%%%%%
%%%%%%%%%%%%%%%%%%%%%%%%%%%%%%%%%%%%%%%%%%%%%%%%%%%%%%%%%%%%%%%%%%%%%%%%%%%%%%%%%%%%%%%
%%%%%%%%%%%%%%%%%%%%%%%%%%%%%%%%%%%%%%%%%%%%%%%%%%%%%%%%%%%%%%%%%%%%%%%%%%%%%%%%%%%%%%%
%%%%%%%%%%%%%%%%%%%%%%%%%%%%%%%%%%%%%%%%%%%%%%%%%%%%%%%%%%%%%%%%%%%%%%%%%%%%%%%%%%%%%%%
%%%%%%%%%%%%%%%%%%%%%%%%%%%%%%%%%%%%%%%%%%%%%%%%%%%%%%%%%%%%%%%%%%%%%%%%%%%%%%%%%%%%%%%
\section{Discussion and ancillary results}\label{s:DA}
The main results of this work have now been presented. In this section we offer some discussion on several disparate topics relating to interpretation and application of these results. 

\subsection{The physical interpretation of $\left\langle g\right\rangle_{0}$, $\left\langle g\right\rangle_{\emptyset}$ and $\left\langle g\right\rangle_{m}$}\label{s:5}

In the preceding sections we have introduced three different co-moments of the NSF that can be related to moments of the normalized TOF signal. These are $\left\langle g\right\rangle_{0}$, defined by \eqref{e:1.4.4}, $\left\langle g\right\rangle_{\emptyset}$, defined by \eqref{e:3.5b}, and $\left\langle g\right\rangle_{m}$, defined by \eqref{e:4.2}. The three quantities can all be used to infer physical quantities from a TOF signal, but have subtle differences that we discuss here.

First, $\left\langle g\right\rangle_{0}$ represents a \textit{directed yield-averaged co-moment} of the NSF, where the function $g$ identifies a unique such co-moment. The normalization quantity in $\left\langle g\right\rangle_{0}$ is $\langle 1 \rangle$, which represents the neutron yield per unit solid angle emitted in a particular direction $\hat{v}$. Therefore, $\left\langle g\right\rangle_{0}$ represents a weighting of the quantity $g$ by the number of neutrons emitted in the direction defined by $\hat{v}$. It is defined for arbitrary ion distribution functions $F\left(t,\vec{v_{c}},\vec{v_{r}}\right)$.

Secondly, $\left\langle g\right\rangle_{\emptyset}$ was introduced as an approximation to $\left\langle g\right\rangle_{0}$ obtained from expansion of the Jacobian determinant term in $R_{k}$. We also used an expansion of $\left\langle 1\right\rangle^{-1}$ in the computation of $\left\langle g\right\rangle_{\emptyset}$. These expansions were possible due to assumptions about the smallness of $\alpha$, $\beta$ and $\zeta$. These assumptions place restrictions on the ion distribution functions $F\left(t,\vec{v_{c}},\vec{v_{r}}\right)$ for which $\left\langle g\right\rangle_{0}$ has a physical interpretation (effectively, we require $v_{c}$ and $v_{r}$ to be small relative to $v_{0}$). Although the Jacobian determinant has been eliminated in the definition of $\left\langle g\right\rangle_{\emptyset}$, this co-moment can still depend on the direction of neutron emission $\hat{v}$ through the differential cross-section term $\frac{d\sigma}{d\Omega_{cm}}$, see \eqref{e:3.5a}. However, this dependence on direction will be weaker than for $\left\langle g\right\rangle_{}$ and so we can assume that $\left\langle g\right\rangle_{\emptyset}$ is an \textit{approximation for the yield-averaged co-moment}. The approximation will be valid if the differential cross section is close to isotropic. Since we are integrating over all of the ion velocity space $d^{3}\vec{v_{c}}d^{3}\vec{v_{r}}$, anisotropy of $F\left(t,\vec{v_{c}},\vec{v_{r}}\right)$ is unimportant. However, if the ion distribution functions are isotropic then the approximation is exact, regardless of anisotropy of the differential cross section. This is because we can make use of the trivial integration $\int\frac{d\sigma}{d\Omega_{cm}}d\Omega_{r} = \sigma$ which arises when ion distribution functions are isotropic. In fact, this trivial integral arises not just for isotropic ion distributions but simply for those which are independent of the direction of $\vec{v_{r}}$, that is $F = F\left(t,\vec{v_{c}},v_{r}\right)$.

Finally, we come to $\left\langle g\right\rangle_{m}$. The definition of this quantity \eqref{e:4.2} resulted from computation of \eqref{e:3.5b} for the specific case of Maxwellian ion distribution functions. Our choice of ion distribution function allowed co-moments in terms of $\alpha$, $\beta$ and $\zeta$ to be redefined as co-moments of ion temperature $T$ and bulk fluid velocity $\vec{v_{f}}$, as given by \eqref{e:4.3a}-\eqref{e:4.3e}. Furthermore, for Maxwellians the trivial integration over the differential cross section discussed in the last paragraph can be applied and so $\left\langle g\right\rangle_{m}$ represents the \textit{yield-averaged co-moment for a Maxwellian plasma} undergoing thermonuclear burn.

%%%%%%%%%%%%%%%%%%%%%%%%%%%%%%%%%%%%%%%%%%%%%%%%%%%%%%%%%%%%%%%%%%%%%%%%%%%%%%%%%%%%%%%
%%%%%%%%%%%%%%%%%%%%%%%%%%%%%%%%%%%%%%%%%%%%%%%%%%%%%%%%%%%%%%%%%%%%%%%%%%%%%%%%%%%%%%%
%%%%%%%%%%%%%%%%%%%%%%%%%%%%%%%%%%%%%%%%%%%%%%%%%%%%%%%%%%%%%%%%%%%%%%%%%%%%%%%%%%%%%%%
%%%%%%%%%%%%%%%%%%%%%%%%%%%%%%%%%%%%%%%%%%%%%%%%%%%%%%%%%%%%%%%%%%%%%%%%%%%%%%%%%%%%%%%
%%%%%%%%%%%%%%%%%%%%%%%%%%%%%%%%%%%%%%%%%%%%%%%%%%%%%%%%%%%%%%%%%%%%%%%%%%%%%%%%%%%%%%%

\subsection{Variations with neutron emission direction and the case for antipodal detectors}\label{s:7}
Throughout this work we have emphasized the need for detectors to be collinear since the NSF can vary with emission direction. \textit{It remains an open question of how detectors at different distances but located in different neutron emission directions could be used to obtain information about the time-varying conditions of the source.} The collinearity requirement is due to the presence of the $\beta$ term in the expressions for inverse-velocity co-moments. It means that these co-moments will depend on the direction of neutron emission. The $\alpha$ and $\zeta$ terms are independent of neutron emission direction and so will make equal contributions to co-moments in all directions. The moments method that we have outlined in this work can be applied to multiple neutron emission directions, but each direction will require, in principle, $n$ detectors to obtain an independent solution for the co-moments in that direction. However, from \eqref{e:3.10} we see that co-moments of the form $C\left(t,v^{-1}\right)^{j,0}_{0}$ will have a leading order term that is independent of $\beta$ (the leading order term of all other co-moments will have a dependence on $\beta$). Therefore, to leading order, co-moments $C\left(t,v^{-1}\right)^{j,0}_{0}$ will have the same value for all emission directions. With this approximation, the co-moments in multiple emission directions up to order $n-1$ can be found using $n$ detectors in one direction and $n-1$ detectors on each of the other directions.

In the remainder of this section we highlight the unique capability of an antipodal detector configuration. Consider $n$ detectors located in a particular emission direction, $\hat{v}$, and $n$ detectors in the antipodal direction, $-\hat{v}$. We can find the inverse-velocity co-moments up to order $n-1$ for these two emission directions. Now, any terms in these two sets of co-moments that depend on $\beta^{i}$ will be equal when $i$ is even and equal in magnitude but with opposite sign when $i$ is odd. For example, $C\left(t^{p},\beta\right)_{\emptyset}^{1,1}$, $C\left(t,\beta\right)_{\emptyset}^{1,1}$ and $\left\langle\beta\right\rangle_{\emptyset}$, which are the terms of order $v_{0}^{-1}$ in \eqref{e:3.10}, will all be of equal magnitude and opposite sign for neutrons emitted in antipodal directions. Therefore, a summation of the inverse-velocity co-moments inferred from antipodal detectors will eliminate these terms allowing for a more accurate inference of higher order terms in \eqref{e:3.10}. 

This capability of the antipodal detector configuration holds regardless of the form of the ion distribution functions or their time-varying behaviour. It is generally not possible to eliminate lower order terms in this way with other configurations of detectors. One commonly employed exception to this is time-integrated co-moments for Mawellians. The term $\left\langle \vec{v_{f}}\cdot\hat{v}\right\rangle_{m}$ in \eqref{e:4.6b} represents a vector component and so with measurements of this in three arbitrary neutron emission directions we can recover the bulk fluid velocity vector $\left\langle \vec{v_{f}}\right\rangle_{m}$. However, it is important to note that the corresponding time-varying terms, for example $C\left(t,\vec{v_{f}}\cdot\hat{v} \right)^{1,1}_{m}$, are not vector components and so the same treatment cannot be applied.

%%%%%%%%%%%%%%%%%%%%%%%%%%%%%%%%%%%%%%%%%%%%%%%%%%%%%%%%%%%%%%%%%%%%%%%%%%%%%%%%%%%%%%%
%%%%%%%%%%%%%%%%%%%%%%%%%%%%%%%%%%%%%%%%%%%%%%%%%%%%%%%%%%%%%%%%%%%%%%%%%%%%%%%%%%%%%%%
%%%%%%%%%%%%%%%%%%%%%%%%%%%%%%%%%%%%%%%%%%%%%%%%%%%%%%%%%%%%%%%%%%%%%%%%%%%%%%%%%%%%%%%
%%%%%%%%%%%%%%%%%%%%%%%%%%%%%%%%%%%%%%%%%%%%%%%%%%%%%%%%%%%%%%%%%%%%%%%%%%%%%%%%%%%%%%%
%%%%%%%%%%%%%%%%%%%%%%%%%%%%%%%%%%%%%%%%%%%%%%%%%%%%%%%%%%%%%%%%%%%%%%%%%%%%%%%%%%%%%%%

\subsection{Determining a set of suitable distances for detectors}\label{s:A2}
In section \ref{s:2} we learned that having $n$ collinear detectors at different distances from the neutron source will allow inverse-velocity co-moments up to order $n-1$ to be calculated by taking central moments of the TOF signal on every detector. Until now, we have not considered if there is an optimum set of distances for these $n$ detectors and what those distances might be. For a given experimental facility there will be many practical considerations for identifying these distances, such as the background scattering environment and detector response. We do not examine these issues here, but instead we determine a set of distances based on how curves of projection bisect the $\left(t,v\right)$ space. 

To begin, we recall that a curve of projection is defined by the parameters $\left(x_{0},\tau_{0}\right)$ and the equation  
\begin{equation}
    \tau_{0} = t+\frac{x_{0}}{v},\label{e:A2.1}
\end{equation} 
where $t$ and $v$ are independent variables. As $x_{0}\rightarrow 0$ curves of projection become orthogonal to the $t$ axis and so each curve of projection includes all neutrons produced at a given time regardless of velocity. In contrast, as $x_{0}\rightarrow\infty$, curves of projection become orthogonal to the $v$ axis and so each curve of projection includes all neutrons produced with a given velocity regardless of production time. For intermediate values of $x$, each curve of projection will be non-orthogonal to both axes and depend on both the time and velocity behaviour of $f\left(t,v\right)$. Intuitively, if we have $n$ detectors then a suitable set of distances will be that for which the curves of projection have as varied a dependence on the time and velocity behaviour of $f\left(t,v\right)$ as possible. This is the principle on which we calculate the set of distances. Given this principle, distances will depend on the shape of $f\left(t,v\right)$. To account for this, we assume that $f\left(t,v\right)$ is well described by values of mean $\left(\mu_{t},\mu_{v}\right)$ and standard deviation $\left(\sigma_{t},\sigma_{v}\right)$, and the variation of $f\left(t,v\right)$ with $t$ and $v$ is uncorrelated.

\begin{figure}
    \centering
    \includegraphics[width=0.9\linewidth]{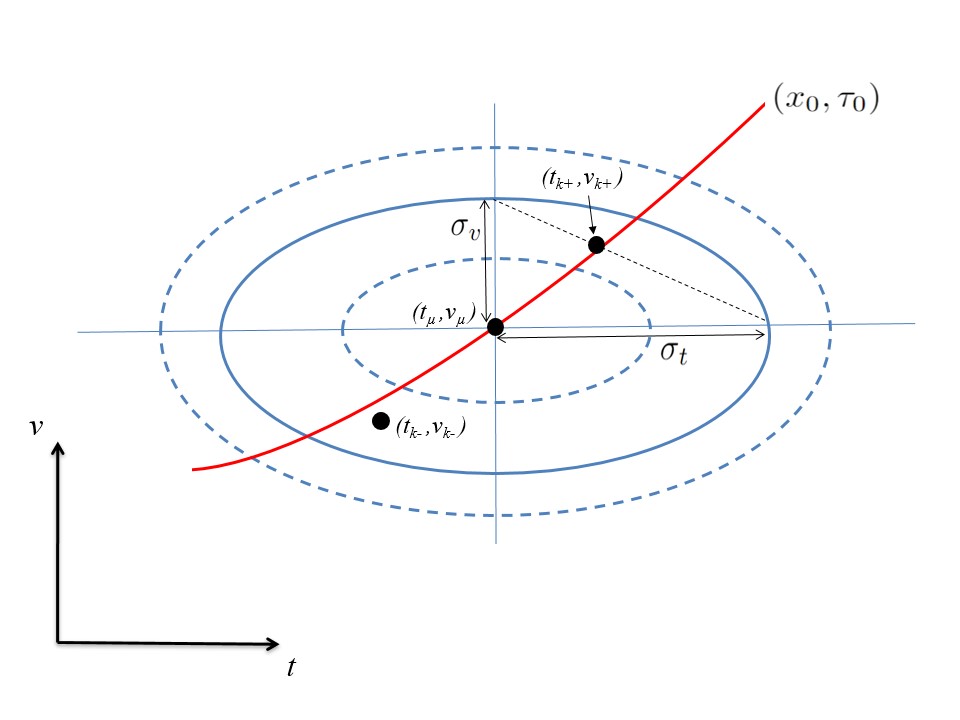}
    \caption{An illustration of $\left(t,v\right)$ space and our strategy for finding the set of suitable detector distances. Blue curves indicate contours of $f\left(t,v\right)$ and black dots indicate a triple defined by \eqref{e:A2.6a}-\eqref{e:A2.6c}. The red curve indicates the best fit curve of projection for this triple and $x_{0}$ will be a suitable distance for locating a detector.}
    \label{fig:n1}
\end{figure}

Given these assumptions for $f\left(t,v\right)$, and $n$ detectors, we define $n$ triples of points in the $\left(t,v\right)$ space, $\left[\left(t_{k-},v_{k-}\right),\left(t_{\mu},v_{\mu}\right),\left(t_{k+},v_{k+}\right)\right]$. The triples are defined by
\begin{eqnarray}
   \left(t_{k-},v_{k-}\right) &=& \left(\mu_{t}-\frac{k}{n-1}\sigma_{t},\mu_{v}-\frac{n-1-k}{n-1}\sigma_{v}\right),\label{e:A2.6a}\\
   \left(t_{\mu},v_{\mu}\right) &=& \left(\mu_{t},\mu_{v}\right),\label{e:A2.6b}\\
   \left(t_{k+},v_{k+}\right) &=& \left(\mu_{t}+\frac{k}{n-1}\sigma_{t},\mu_{v}+\frac{n-1-k}{n-1}\sigma_{v}\right),\label{e:A2.6c}
\end{eqnarray}
for $k=0,\ldots,n-1$ and $n>1$. One such triple is illustrated in fig. \ref{fig:n1}. For each triple we can now ask what curve of projection is the best fit, in a least squares sense, to the three points of the triple. Using \eqref{e:A2.1} this curve of projection $\left(x_{0},\tau_{0}\right)$ will be the solution to 
\begin{equation}
    \begin{bmatrix}
-v_{k-}^{-1} & 1\\
-v_{\mu}^{-1} & 1\\
-v_{k+}^{-1} & 1
\end{bmatrix}\begin{bmatrix}
x_{0} \\
\tau_{0} 
\end{bmatrix} = \begin{bmatrix}
t_{k-}\\
t_{\mu}\\
t_{k+}
\end{bmatrix},\nonumber\label{e:A2.3}
\end{equation}
which is of the form
\begin{equation}
    \mathbf{A}\mathbf{x} = \mathbf{b},\nonumber\label{e:A2.4}
\end{equation}
and can be solved for $\mathbf{x}$ using 
\begin{equation}
    \mathbf{x} = \left(\mathbf{A}^{T}\mathbf{A}\right)^{-1}\mathbf{A}^{T}\mathbf{b}.\nonumber\label{e:A2.5}
\end{equation}
The value of $x_{0}$ contained $\mathbf{x}$ will be the distance to a detector such that neutrons produced at $\left(t_{k-},v_{k-}\right)$, $\left(t_{\mu},v_{\mu}\right)$ and $\left(t_{k+},v_{k+}\right)$ all arrive at the detector as close to time $\tau_{0}$ as possible. The triples defined by \eqref{e:A2.6a}-\eqref{e:A2.6c} will result in $x_{0} = 0$ and $x_{0} = \infty$ for $k=0$ and $k=n-1$, respectively, in agreement with our intuition. 
This solution is particularly useful as it illustrates how the detector locations will vary with parameters such as $\mu_{v}$ and $\sigma_{v}$, which can, for example, be estimated from the nuclear reaction (DD or DT) and ion temperature, respectively, and $\sigma_{t}$ which can be estimated from thermonuclear burn duration. Therefore, the detector distances are not fixed but will depend on the properties of the experiment under investigation. Nevertheless, it is likely that having a detector as close to, and as far from, the neutron source as is feasible will always be desirable.

\subsection{Generalized moments in the velocity domain}\label{s:6}
In section \ref{s:0} we described how the moments method has been rigorously investigated in the neutron velocity domain. Recent work\cite{APPELBE_HEDP2024} has demonstrated that a set of "generalized moments" exist that can suppress contributions of the $\vec{v_{c}}\cdot\hat{v}$ term to the shape of the neutron spectrum. For Maxwellians, this suppresses the contributions of bulk fluid velocity which could help obtain a more accurate measurement of ion temperature or identification of non-Maxwellian ion distributions. In the present work we chose, for simplicity, to work with central moments in the time-of-flight domain. We have not attempted to investigate if a set of functions $g$ exist which could provide additional information on the ion distribution functions. \textit{It remains an open question if there exists a set of generalized moments in the time-of-flight domain which could suppress the effects of the $\vec{v_{c}}\cdot\hat{v}$ term for a time-dependent neutron source.}

In this section, we consider the application of the generalized moments in the velocity domain to TOF signals. To do this, we first note that \eqref{e:1.1} defines the relationship between the time-of-flight domain and the velocity domain. From this we can write
\begin{eqnarray}
    v &=& \frac{x}{\tau}\left(1+\epsilon\right),\label{e:6.4c}\\
    \epsilon &=& \frac{vt}{x}.\label{e:6.4d}
\end{eqnarray}
The parameter $\epsilon$ obeys the following limits
\begin{eqnarray}
   \epsilon &\rightarrow& 0 \quad\text{as}\quad x\rightarrow\infty,\label{e:6.4a}\\
   \epsilon^{-1} &\rightarrow& 0  \quad\text{as}\quad x\rightarrow 0.\label{e:6.4b}
\end{eqnarray}
Equations $17$ and $18$ of Appelbe et al\cite{APPELBE_HEDP2024} define the generalized moments in the velocity domain. We can find the corresponding moments in the time-of-flight domain by replacing $v$ in those equations with $x\tau^{-1}$ and using \eqref{e:6.4c}. This results in 
\begin{eqnarray}
g_{n} &=& -\frac{n+\delta_{n}}{2^{n+1}}\left(\frac{x\left(1+\epsilon\right)}{v_{0}\tau}\right)^{\delta_{n}}\left(\left(\frac{x\left(1+\epsilon\right)}{v_{0}\tau}\right)^{2}-1\right)^{n}\nonumber\\
&&\times\left(\left(\frac{x\left(1+\epsilon\right)}{v_{0}\tau}\right)^{2}-1-\frac{2}{n+\delta_{n}}\right),\label{e:6.6a}\\
\langle g_{n} \rangle_{-2} &=& \frac{\langle\left(\tau x^{-1}\left(1+\epsilon\right)^{-1}\right)^{2}  g_{n} \rangle}{\langle \left(\tau x^{-1}\left(1+\epsilon\right)^{-1}\right)^{2} \rangle}.\label{e:6.6b}
\end{eqnarray}
where $n\in\mathbb{N}$ and $\delta_{n} = n\,mod\, 2$, i.e. $\delta_{n}$ is $0$ for $n$ even and $1$ for $n$ odd. Note the subscript $-2$, in \eqref{e:6.6b} which denotes a normalization that was chosen to suppress $\vec{v_{c}}\cdot\hat{v}$ contributions.

The utility of \eqref{e:6.6a} and \eqref{e:6.6b} for analysing TOF signals is clear. As $x\rightarrow\infty$, \eqref{e:6.6a} and \eqref{e:6.6b} recover the generalized moments in the velocity domain. As $x$ becomes smaller, the $\left(1+\epsilon\right)$ term acts as a measure of how non-instantaneous neutron production affects the generalized velocity moments. As $x\rightarrow 0$, the utility of the generalized moments breaks down since the TOF signal is dominated by the neutron production time.

This analysis only requires a single detector. Since \eqref{e:6.6a} is not a polynomial in $x$ we cannot use the procedure developed in section \ref{s:2} to obtain separate co-moments from moments of the TOF. Therefore, in addition to the open question of whether a set of generalized moments exist in the time-of-flight domain, \textit{we can also ask how such generalized moments would be applied to multiple detectors at different distances.}

\section{Conclusions}\label{s:C}
We have attempted to lay the theoretical foundations for analyzing time-of-flight signals on multiple collinear detectors using the moments method. Much work remains in building a more practical structure on these foundations, as suggsted by our list of assumptions in section \ref{s:0}. Nevertheless, here we offer some guidance on the utility of our work. 

First, with three collinear detectors we can solve for $C\left(t,v^{-1}\right)^{2,0}_{0}$, the variance of the neutron production time. This could be beneficial for experiments in which knowledge of the neutron pulse duration is important but detectors cannot be placed sufficiently close to the source to measure on a single detector. Secondly, we can also solve for $C\left(t,v^{-1}\right)^{1,1}_{0}$ using three collinear detectors. For thermonuclear plasmas with Maxwellian ion distributions this provides a measure of the acceleration of the neutron emitting plasma (for example, the hotspot acceleration in ICF), while for beam-target plasmas it provides a measure for the rate of change of beam energy with time. Finally, with four collinear detectors we can solve for $C\left(t,v^{-1}\right)^{1,2}_{0}$, which provides information on how ion temperature is changing with time in thermonuclear plasmas. 

In conclusion, we note that each individual inverse-velocity co-moment provides unique information on the neutron emitting plasma. Therefore, increasing the number of co-moments that can be measured and the number of neutron emission directions on which they can be measured will allow for an increasingly detailed characterization of the plasma. With this in mind, we hope that our work will provide motivation for neutron detector improvements and developments at both existing and future facilities.

%%%%%%%%%%%%%%%%%%%%%%%%%%%%%%%%%%%%%%%%%%%%%%%%%%%%%%%%%%%%%%%%%%%%%%%%%%%%%%%%%%%%%%%
%%%%%%%%%%%%%%%%%%%%%%%%%%%%%%%%%%%%%%%%%%%%%%%%%%%%%%%%%%%%%%%%%%%%%%%%%%%%%%%%%%%%%%%
%%%%%%%%%%%%%%%%%%%%%%%%%%%%%%%%%%%%%%%%%%%%%%%%%%%%%%%%%%%%%%%%%%%%%%%%%%%%%%%%%%%%%%%
%%%%%%%%%%%%%%%%%%%%%%%%%%%%%%%%%%%%%%%%%%%%%%%%%%%%%%%%%%%%%%%%%%%%%%%%%%%%%%%%%%%%%%%
%%%%%%%%%%%%%%%%%%%%%%%%%%%%%%%%%%%%%%%%%%%%%%%%%%%%%%%%%%%%%%%%%%%%%%%%%%%%%%%%%%%%%%%

\appendix
\section{Reaction kinematics and differential cross section}\label{s:A1}
In this appendix we give details on some reaction kinematics relations and properties of differential cross sections that we have used in our work.

\subsection{Reaction kinematics}\label{s:A1.1}
We use $v_{0}$ to denote the neutron velocity in the CM frame when $v_{r} = 0$. Since the pair of reactants have zero kinetic energy the neutron velocity is entirely dependent on the mass difference between reaction products and reactants and so will be a constant for any reaction. Its value is given by
\begin{equation}
        v_{0} = \sqrt{\frac{2m_{4}}{m_{n}\left(m_{n}+m_{4}\right)}Q},\nonumber
    \end{equation}
where $Q$ is the Q-value of the reaction, $m_{n}$ is neutron mass and $m_{4}$ is the particle mass of the other reaction product ($He^{3}$ for DD reactions and $He^{4}$ for DT reactions). For DD neutrons, $v_{0}\approx 2.1602\times 10^{7}\,m\,s^{-1}$, and for DT neutrons, $v_{0}=5.1234\times 10^{7}\,m\,s^{-1}$.

When $v_{r}\neq 0$ we use $\vec{u}$ to denote the neutron velocity vector in the CM frame. This vector is related to the neutron velocity vector in the lab frame by $\vec{v} = \vec{v_{c}}+\vec{u}$ and has a magnitude
\begin{equation}
        u = \sqrt{v_{0}^{2}+\nu v_{r}^{2}},\nonumber
\end{equation}
where
\begin{equation}
        \nu = \frac{m_{1}m_{2}m_{4}}{m_{3}\left(m_{1}+m_{2}\right)\left(m_{n}+m_{4}\right)},\label{en:3}
\end{equation}
with $m_{1}$ and $m_{2}$ being the particle masses of the reactants. The vector $\vec{u}$ can be directed in any direction in the CM frame, with a probability that is defined by the differential cross section, which we next outline.

%%%%%%%%%%%%%%%%%%%%%%%%%%%%%%%%%%%%%%%%%%%%%%%%%%%%%%%%%%%%%%%%%%%%%%%%%%%%%%%%%%%%%%%
%%%%%%%%%%%%%%%%%%%%%%%%%%%%%%%%%%%%%%%%%%%%%%%%%%%%%%%%%%%%%%%%%%%%%%%%%%%%%%%%%%%%%%%
%%%%%%%%%%%%%%%%%%%%%%%%%%%%%%%%%%%%%%%%%%%%%%%%%%%%%%%%%%%%%%%%%%%%%%%%%%%%%%%%%%%%%%%

\subsection{Reaction differential cross section}\label{s:A1.2}
The differential cross section is denoted by $\frac{d\sigma}{d\Omega_{cm}}d\Omega_{cm}$. It represents the probability of two particles undergoing a reaction and emitting a neutron in a given direction in the CM frame, and has units of per unit area per unit solid angle. It is a function of $v_{r}$ (or, equivalently, the reaction energy, $\frac{1}{2}m_{12}v_{r}^{2}$) and a scattering angle, which is specified by the direction of neutron emission in the CM frame. The scattering angle is defined by $\hat{v_{r}}\cdot\hat{u}$, where $\hat{v_{r}}$ and $\hat{u}$ are unit vectors in the direction of $\vec{v_{r}}$ and $\vec{u}$, respectively. Given this definition of the scattering angle, we have
\begin{equation}
    \int \frac{d\sigma}{d\Omega_{cm}}d\Omega_{cm} = \int \frac{d\sigma}{d\Omega_{cm}}d\Omega_{r} = \sigma,\nonumber
\end{equation}
where subscript $r$ denotes integration over the solid angle of $\hat{v_{r}}$, and subscript $cm$ denotes integration over the solid angle of $\hat{u}$.

The differential cross section can be transformed to be a function of neutron emission direction in the lab frame (defined by $\hat{v}$) using the Jacobian determinant for a solid angle transformation\cite{Catchen_JCP1978}
\begin{equation}
\frac{d\Omega}{d\Omega_{cm}} = \frac{u^{2}}{v^{2}}\cos\delta\nonumber
\end{equation}
where $\delta$ is the angle between $\hat{u}$ and $\hat{v}$ ($\cos\delta = \hat{u}\cdot\hat{v}$), and $d\Omega$ indicates the solid angle of $\hat{v}$. This transformation results in
\begin{equation}
\frac{d\sigma}{d\Omega_{cm}}d\Omega_{cm} = \frac{v^{2}}{u\sqrt{u^{2}-v_{c}^{2}+\left(\vec{v_{c}}\cdot\hat{v}\right)^{2}}}\frac{d\sigma}{d\Omega_{cm}}d\Omega,\nonumber
\end{equation}
where the scattering angle is now given by
\begin{equation}
\hat{v_{r}}\cdot\hat{u} = \frac{\hat{v_{r}}\cdot\left(\vec{v}-\vec{v_{c}}\right)}{u}.\nonumber
\end{equation}
Finally, an isotropic cross section is one which depends on $v_{r}$ only (it is independent of scattering angle) and so 
\begin{equation}
\frac{d\sigma}{d\Omega_{cm}}  = \frac{\sigma}{4\pi},\nonumber
\end{equation}
where $\sigma$ is the reaction cross section. We note that this definition implies isotropy in the CM frame only and the above expression for transformation into the lab frame is still required.

%%%%%%%%%%%%%%%%%%%%%%%%%%%%%%%%%%%%%%%%%%%%%%%%%%%%%%%%%%%%%%%%%%%%%%%%%%%%%%%%%%%%%%%
%%%%%%%%%%%%%%%%%%%%%%%%%%%%%%%%%%%%%%%%%%%%%%%%%%%%%%%%%%%%%%%%%%%%%%%%%%%%%%%%%%%%%%%
%%%%%%%%%%%%%%%%%%%%%%%%%%%%%%%%%%%%%%%%%%%%%%%%%%%%%%%%%%%%%%%%%%%%%%%%%%%%%%%%%%%%%%%
%%%%%%%%%%%%%%%%%%%%%%%%%%%%%%%%%%%%%%%%%%%%%%%%%%%%%%%%%%%%%%%%%%%%%%%%%%%%%%%%%%%%%%%
%%%%%%%%%%%%%%%%%%%%%%%%%%%%%%%%%%%%%%%%%%%%%%%%%%%%%%%%%%%%%%%%%%%%%%%%%%%%%%%%%%%%%%%

\section{Identities used in the expansion of inverse-velocity co-moments}\label{s:A5}
In this appendix we list two identities that are used in section \ref{s:3} for obtaining approximations to the inverse-velocity co-moments.

\subsection{The multinomial theorem}\label{s:A5.1}
The multinomial theorem is a generalization of the binomial theorem that describes how to expand a power of a sum of an arbitrary number of terms. Given $m$ terms, $x_{1},x_{2},\ldots,x_{m}$, the multinomial theorem is defined as
\begin{equation}\label{e:A5.1}
    \left(x_{1}+\ldots+x_{m}\right)^{n} = \sum_{k_{1}+\ldots+k_{m}=n}\binom{n}{k_{1},\ldots,k_{m}}\prod_{r=1}^{m}x_{r}^{k_{r}},\nonumber
\end{equation}
where
\begin{equation}\label{e:A5.2}
    \binom{n}{k_{1},\ldots,k_{m}}= \frac{n!}{k_{1}!\ldots k_{m}!},\quad k_{1},\ldots,k_{m}\in\mathbb{N}_{0},\nonumber
\end{equation}
and the summation index indicates that the sum is taken over all combinations of indices $k_{1}$ through $k_{m}$ such that the sum of these indices is $n$. The theorem can be applied to an infinite series if the series converges absolutely. In section \ref{s:3}, our assumption about the smallness of of $\beta$, $\alpha$, $\zeta$, \eqref{e:3.3} allows us to assume such convergence. 

\subsection{Alternating sums of binomial coefficients}\label{s:A5.2}
Here we prove an identity used in section \ref{s:3.4}. The identity is 
\begin{equation}\label{e:A5.4}
    \sum_{k=0}^{n}\binom{n}{k}\left(-1\right)^{k}k^{m} = 0,\quad m<n,\quad m\in\mathbb{N}_{0},\nonumber
\end{equation}
where
\begin{equation}\label{e:A5.3}
    \binom{n}{k} = \frac{n!}{k!\left(n-k\right)!},\nonumber
\end{equation}
is the binomial coefficient.

To prove, we start with the case of $m=0$, for which we have
\begin{eqnarray}\label{e:A5.5}
     \sum_{k=0}^{n}\binom{n}{k}\left(-1\right)^{k} &=&  \sum_{k=0}^{n}\binom{n}{k}\left(-1\right)^{k}\left(1\right)^{n-k}\nonumber\\
     &=& \left(1+\left(-1\right)\right)^{n} = 0.
\end{eqnarray}
Next, for $m>0$ we have
\begin{eqnarray}
   \sum_{k=0}^{n}\binom{n}{k}\left(-1\right)^{k}k^{m} &&=  \sum_{k=1}^{n}\binom{n-1}{k-1}\left(-1\right)^{k}nk^{m-1},\nonumber\\
   &&=  n\sum_{k=0}^{n-1}\binom{n-1}{k}\left(-1\right)^{k+1}\left(k+1\right)^{m-1}\nonumber\\
   =  -n\sum_{k=0}^{n-1}\binom{n-1}{k}&&\left(-1\right)^{k}\left[\sum_{j=0}^{m-1}\binom{m-1}{j}k^{j}\right],\label{e:A5.6}
\end{eqnarray}
We can continue to recursively apply \eqref{e:A5.6}, reducing the exponent on $k$ with each recursion until we have a sum of terms of the form \eqref{e:A5.5}, which, as we have shown, is equal to $0$.

%%%%%%%%%%%%%%%%%%%%%%%%%%%%%%%%%%%%%%%%%%%%%%%%%%%%%%%%%%%%%%%%%%%%%%%%%%%%%%%%%%%%%%%
%%%%%%%%%%%%%%%%%%%%%%%%%%%%%%%%%%%%%%%%%%%%%%%%%%%%%%%%%%%%%%%%%%%%%%%%%%%%%%%%%%%%%%%
%%%%%%%%%%%%%%%%%%%%%%%%%%%%%%%%%%%%%%%%%%%%%%%%%%%%%%%%%%%%%%%%%%%%%%%%%%%%%%%%%%%%%%%
%%%%%%%%%%%%%%%%%%%%%%%%%%%%%%%%%%%%%%%%%%%%%%%%%%%%%%%%%%%%%%%%%%%%%%%%%%%%%%%%%%%%%%%
%%%%%%%%%%%%%%%%%%%%%%%%%%%%%%%%%%%%%%%%%%%%%%%%%%%%%%%%%%%%%%%%%%%%%%%%%%%%%%%%%%%%%%%

\section{Covariance and rates of change}\label{s:A4}
Expressions of the form $C\left(t,b\right)^{1,1}_{0}$ and $C\left(t,b\right)^{1,1}_{m}$ appear frequently in our results in sections \ref{s:3} and \ref{s:4}. For example, in the Maxwellian expressions obtained in section \ref{s:4}, $b$ represents $T$, $\vec{v_{f}}\cdot\hat{v}$, $\left(\vec{v_{f}}\cdot\hat{v}\right)^{2}$, $v_{f}^{2}$, $v_{r}^{2}$, etc. We refer to these expressions as the \textit{covariance} of $b$ with $t$. In this appendix we show that the covariance is a measure of $\frac{\partial b}{\partial t}$, the rate of change of $b$ with respect to $t$, and how we can estimate $\frac{\partial b}{\partial t}$ from measurements of covariance terms. 

We start by \textit{assuming that the acceleration, $\frac{\partial b}{\partial t}$, is constant}, given by  
\begin{equation}\label{e:A4.1}
    b = \frac{\partial b}{\partial t}\left(t-t_{1}\right)+b_{1},
\end{equation}
where $b_{1}$ is the value of $b$ at some time $t_{1}$. Clearly, $b$ in \eqref{e:A4.1} is a function of time $t$. We can think of $b$ as the directed yield-averaged value of $b$ at time $t$, that is, it is averaged over all neutrons produced instantaneously at $t$. Using \eqref{e:A4.1} we have
\begin{equation}\label{e:A4.2}
    \left\langle bt\right\rangle_{0} = \frac{\partial b}{\partial t}\left\langle\left(t-t_{1}\right)t\right\rangle_{0}+b_{1}\left\langle t\right\rangle_{0}.
\end{equation}

Next, we assume $b_{1}=\left\langle b\right\rangle_{0}$ and $t_{1}=\left\langle t\right\rangle_{0}$. The physical \textit{assumption is that the directed yield-averaged value of $b$ for neutrons produced instantaneously at the time $\left\langle t\right\rangle_{0}$ is equal to the directed yield-averaged co-moment of $b$, $\left\langle b\right\rangle_{0}$}. This assumption is reasonable for NSF such as those produced in ICF experiments. With this assumption \eqref{e:A4.2} becomes
\begin{equation}\label{e:A4.3}
    \left\langle bt\right\rangle_{0}-\left\langle b\right\rangle_{0}\left\langle t\right\rangle_{0} = \frac{\partial b}{\partial t}\left\langle\left(t-\left\langle t\right\rangle_{0}\right)t\right\rangle_{0}.
\end{equation}
Finally, since $\left\langle\left(t-\left\langle t\right\rangle_{0}\right)t\right\rangle_{0}=\left\langle\left(t-\left\langle t\right\rangle_{0}\right)^{2}\right\rangle_{0} = C\left(t,b\right)^{2,0}_{0}$ we have
\begin{equation}\label{e:A4.4}
    \frac{\partial b}{\partial t} = \frac{C\left(t,b\right)^{1,1}_{0}}{C\left(t,b\right)^{2,0}_{0}}.
\end{equation}
This equation shows that a covariance of a quantity with time divided by the second order velocity-integrated co-moment (colloquially, the variance of the burn history) gives the rate of change with time of that quantity. When the assumption about constant acceleration does not hold, the right-hand side of \eqref{e:A4.4} can be interpreted as a measure of the average rate of change with time of the quantity $b$ during the neutron emission. 

The two assumptions made above also allow the result of \eqref{e:A4.4} to be generalized as follows
\begin{equation}\label{e:A4.5}
    \left(\frac{\partial b}{\partial t}\right)^{k} =  \frac{C\left(t,b\right)^{j,k}_{0}}{C\left(t,b\right)^{j+k,0}_{0}}.
\end{equation}
Thus, the quantity $\frac{\partial b}{\partial t}$ can be calculated from inverse-velocity co-moments of any order $\geq 2$. Comparing the values obtained from different orders may provide a way of quantify the validity of the assumptions used to obtain \eqref{e:A4.4} and \eqref{e:A4.5}.

\bibliographystyle{unsrt}
\bibliography{Cherwell_bib}

\end{document}